%% file: THz_ADC_arXiv.tex
\documentclass[journal]{IEEEtran}
\usepackage{cite}
\usepackage{amsmath,amssymb,amsfonts}
\usepackage{algorithmic}
\usepackage{graphicx}
\usepackage{textcomp}

\usepackage[caption=false,font=footnotesize]{subfig}
\usepackage{stfloats}
\usepackage{url}
\usepackage{verbatim}
\usepackage{graphicx}
\usepackage{cite}
\usepackage{mathtools}
\usepackage[nolist]{acronym}
\usepackage{siunitx}
\usepackage[hidelinks]{hyperref}
\usepackage{bm}

\def\BibTeX{{\rm B\kern-.05em{\sc i\kern-.025em b}\kern-.08em
    T\kern-.1667em\lower.7ex\hbox{E}\kern-.125emX}}

\graphicspath{{figures/}}

\newcommand\plotwidth{0.45}
\newcommand\plotwidthh{0.22}

\input{abbreviations}
\input{macros}

\begin{document}
\title{Detection Schemes with Low-Resolution \acsp*{ADC} and Spatial Oversampling for Transmission with Higher-Order Constellations in the \acl{THz} Band\\
\thanks{This article was presented in part at the 2022 IEEE Latin-American Conference on Communications (LATINCOM)~\cite{Forsch2022}. This work was supported by a gift from Qualcomm Technologies, Inc.}
\thanks{Christian Forsch and Wolfgang Gerstacker are with the Institute for Digital Communications, Friedrich-Alexander-Universität Erlangen-Nürnberg, Erlangen, Germany (e-mail: christian.forsch@fau.de; wolfgang.gerstacker@fau.de).}
\thanks{Peter Zillmann, Osama Alrabadi, and Stefan Brueck are with Qualcomm CDMA Technologies, Nürnberg, Germany (e-mail: pzillman@qti.qualcomm.com; osamaa@qti.qualcomm.com; sbrueck@qti.qualcomm.com).}
}

\author{Christian~Forsch,~\IEEEmembership{Graduate Student Member,~IEEE}, Peter~Zillmann, Osama~Alrabadi, Stefan~Brueck, and~Wolfgang~Gerstacker,~\IEEEmembership{Senior Member,~IEEE}}

\markboth
{Forsch \MakeLowercase{\textit{et al.}}: Detection Schemes with Low-Resolution ADCs and Oversampling in the THz Band}
{Forsch \MakeLowercase{\textit{et al.}}: Detection Schemes with Low-Resolution ADCs and Oversampling in the THz Band}

\maketitle

\begin{abstract}
In this work, we consider \ac{THz} communications with low-resolution uniform quantization and spatial oversampling at the receiver side, corresponding to a single-input multiple-output (SIMO) transmission.\acused{SIMO}
We fairly compare different \ac{ADC} parametrizations by keeping the \ac{ADC} power consumption constant.
Here, 1-, 2-, and 3-bit quantization is investigated with different oversampling factors.
We analytically compute the statistics of the detection variable, and we propose the optimal and several suboptimal detection schemes for arbitrary quantization resolutions.
Then, we evaluate the \ac{SER} of the different detectors for 16- and 64-ary \ac{QAM}.
The results indicate that there is a noticeable performance degradation of the suboptimal detectors compared to the optimal detector when the constellation size is larger than the number of quantization levels.
Furthermore, at low \acp{SNR}, 1-bit quantization outperforms 2- and 3-bit quantization, respectively, even when employing higher-order constellations.
We confirm our analytical results by Monte Carlo simulations.
Both a pure \ac{LoS} and a more realistically modeled indoor \ac{THz} channel are considered.
Then, we optimize the input signal constellation with respect to \ac{SER} for 1- and 2-bit quantization.
The results give insights for optimizing higher-order constellations for arbitrary quantization resolutions and show that the minimum \ac{SER} can be lowered significantly by appropriately placing the constellation points.
\end{abstract}


\maketitle

\acresetall

\section{Introduction}\label{sec:intro}
\IEEEPARstart{F}{uture} wireless communication systems are expected to provide ultra-high data rates with extremely low latency in order to enable a plethora of new applications~\cite{Akyildiz2020}.
The large bandwidths and high symbol rates which are required for such applications and which can be realized in the \ac{THz} band necessitate very high sampling frequencies of \acp{ADC}, causing a high power consumption.
This problem can be tackled by decreasing the quantization resolution of the \ac{ADC}~\cite{Murmann2015}, arriving at low-resolution quantization such as 1-, 2-, or 3-bit quantization.
However, such quantization with only few bit limits the spectral efficiency to \SIrange{1}{3}{\textnormal{bit per channel use}} (\acsu{bpcu}) per real dimension for Nyquist-rate sampling~\cite{Singh2009}.
Oversampling provides a remedy to increase the achievable rate by retrieving some of the information lost due to quantization~\cite{Gilbert1993,Shamai1994,Koch2010,Zhang2012}.
This can be achieved in the temporal domain by increasing the sampling frequency or in the spatial domain by receiving the useful signal via multiple antennas/channels.
Therefore, in this work, we consider the case of low-resolution quantization with spatial oversampling at the receiver side.
Furthermore, we focus on higher-order constellations, implying that the number of quantization intervals of the \ac{ADC} is typically smaller than the number of possible transmit symbols in our scenarios.

Communication systems with oversampled low-resolution quantization have been already previously studied in the literature.
In particular, the case of 1-bit quantization was considered frequently in previous works due to its relevance and simplicity.
In~\cite{Dabeer2006}, a spread spectrum system with 1-bit quantization was considered which has similarities to our system model.
It was shown that binary \ac{PSK} is not optimal anymore with respect to the channel capacity when observing the transmitted symbol multiple times. 
In~\cite{Krone2010} and~\cite{Krone2012}, it was demonstrated that oversampling can increase the capacity of communication systems with 1-bit quantization to more than \SI{1}{\acs{bpcu}} per real dimension.
In~\cite{Krone2012a}, the results of~\cite{Krone2010} and~\cite{Krone2012} were extended to a more realistic system model with \ac{ISI} which is even beneficial for transmitting with higher-order constellations, especially at high \acp{SNR}.
In~\cite{Landau2018}, the achievable rate for different modulation schemes including higher-order modulation was investigated and the benefit of oversampled 1-bit quantization was confirmed.
The authors of~\cite{Jacobsson2015} showed that higher-order constellations can be detected with 1-bit quantizers in a spatially oversampled system in case the number of receive antennas is sufficiently high and the \ac{SNR} is appropriate.
Thus, these previous works motivate the usage of higher-order constellations for 1-bit quantization.

Furthermore, there are also results on multi-bit quantization.
In\cite{Zhang2016}, the achievable rate of massive \ac{MIMO} systems, which correspond to spatial oversampling, with 1- and 2-bit quantization was analyzed.
The results show that the use of higher-order constellations is feasible.
In~\cite{Jacobsson2017}, it was also demonstrated that the achievable rate can be increased for 2-bit quantization with higher-order constellations such as 64-ary \ac{QAM}, and it was shown that the performance in terms of achievable rate for 3-bit quantization is very close to the unquantized case.

In addition to the achievable rate analysis of low-resolution quantization with oversampling, which was mainly conducted in the above referenced works, the performance in terms of error rate for higher-order constellations is of interest.
In~\cite{Halsig2014}, the \ac{SER} for different symbol sources was analyzed for 1-bit quantization with oversampling.
The results indicate that 4-ary \ac{ASK} with independent and uniformly distributed symbols results in a high error floor for the utilized detector.
In~\cite{Ugurlu2016}, it was shown via numerical evaluation that the \ac{SER} for quantization with 1-3 bit for higher-order constellations can be reduced by inducing artificial noise in a maximum ratio combining receiver.
The performance for \ac{PSK}-modulated symbols was investigated in~\cite{Sun2021} for a massive \ac{MIMO} scenario with 1-bit quantization.
According to the presented \ac{SER} results, the spatial oversampling enables a reliable detection even for higher-order \ac{PSK} constellations.
There are various further works which deal with massive \ac{MIMO} systems with low-resolution quantization and higher-order constellations, e.g.,~\cite{Choi2015,Wang2015,Choi2016,Hong2018,Uecuencue2020}.
In these works, different detection schemes are developed and their viability is shown via Monte Carlo simulations.
Recently, the authors in~\cite{Yilmaz2024,Yilmaz2024a} proposed a pseudo-random quantization scheme for 1-bit massive \ac{MIMO} systems which enhances the performance beyond that for adding random and unknown artificial noise as in~\cite{Ugurlu2016}.
Furthermore, they proposed a two-stage detector for single-carrier multi-user systems assuming frequency-flat channels and a frequency-domain equalization scheme for frequency-selective fading channels with \ac{OFDM} transmission.
The analysis, however, is limited to 1-bit quantization as opposed to our general approach with arbitrary quantization resolution.

The above referenced works do not provide any exact analytical results on the \ac{SER} of higher-order \ac{QAM} or \ac{ASK} constellations.
Some corresponding results were presented in~\cite{Nakashima2017} and~\cite{Nakashima2018}.
Here, the \ac{SER} for 1-bit quantization with different oversampling factors and a 4-\ac{ASK} constellation was analyzed, and the optimal \ac{ML} detector was developed.
However, multi-bit quantization was not studied in~\cite{Nakashima2017} and~\cite{Nakashima2018}.

More recently, low-resolution quantization has been considered particularly for \ac{THz} communications.
In \cite{Neuhaus2020}, a transmission with \ac{ZXM} signals specifically tailored to 1-bit quantization and temporal oversampling at the receiver is studied in the sub-\ac{THz} band.
Finite-state machines are developed for an efficient demodulation of the {\ac{ZXM}} transmit signals.
Similarly as in our work, a single-antenna transmitter and a receiver with multiple antennas are assumed in a \ac{LoS} scenario.
Conventional linear modulation and higher-order constellations are not considered.
In~\cite{He2021}, a deep learning-assisted {\ac{THz}} receiver is designed for a \ac{SISO} transmission with 1-bit quantization, temporal oversampling, and {\ac{THz}} device imperfections.
An optimum {\ac{ML}} detector is derived for quadrature {\ac{PSK}}, and the device imperfections are combatted via a twin-phase training strategy and a neural network-based demodulator.
In \cite{Zhang2022}, a downlink multi-user indoor \ac{THz}
communication system with distance-aware multi-carrier modulation, an array-of-subarrays architecture with hybrid precoding, and low-resolution \acp{DAC} and \acp{ADC} is investigated.
The achievable rate is analyzed, and numerical results indicate that with moderate resolution {\acp{DAC}} and {\acp{ADC}} with 3-5 bit almost the same performance as in the infinite resolution case can be obtained.
No specific modulations are considered, and no error rate analysis is conducted.

In this work, based on~\cite{Forsch2022}, we extend the results in~\cite{Nakashima2017} and~\cite{Nakashima2018} by providing a more detailed analysis on the 1-bit quantization case and considering multi-bit quantization in addition.
The generalization to arbitrary quantization resolution is also an extension of our previous work~\cite{Forsch2022}.
Our analysis leads to exact analytical results on the {\ac{SER}} for quantized reception with arbitrary quantization resolution and {\ac{QAM}} and {\ac{ASK}} constellation sizes, respectively.
Corresponding results have not been available in the literature yet.
We consider a simplified {\ac{THz}} channel model as well as a more realistic indoor {\ac{THz}} channel model and show that the developed optimal symbol detector performs well also for realistically modeled {\ac{THz}} band indoor channels.
We compare the performance for different quantization resolutions with each other while constraining the total \ac{ADC} power consumption to remain constant.
Such approach was also adopted in some existing works, e.g.,~\cite{Uecuencue2020} and~\cite{Cheng2021}, however, not for higher-order constellations.
Moreover, we perform a constellation optimization which was also conducted in~\cite{Singh2009} and~\cite{Krone2012}.
However, our approach is different since we do not focus on capacity-achieving constellations but construct constellations which achieve a minimum \ac{SER} for a given constellation size and \ac{SNR}.
Compared to our previous work~\cite{Forsch2022}, we investigate not only 4-\ac{ASK} but also larger constellations as well as multi-bit quantization.
Our contributions can be summarized as follows:
\begin{itemize}
	\item We derive the optimal \ac{ML} detector for an arbitrary quantization resolution when performing the detection based on an average of the oversampled quantized observations in a frequency-flat single-path \ac{LoS} \ac{THz} channel.
	\item We compare the performance of the optimal \ac{ML} detector to that of two suboptimal detection schemes.
    It can be observed that there is a noticeable performance degradation of the suboptimal detectors when the constellation size is larger than the number of quantization levels of the \ac{ADC}.
	\item We compare the detector performance for quantization with 1-3 bit while keeping the total \ac{ADC} power consumption constant.
	Thereby, we show that 1-bit quantization outperforms 2- and 3-bit quantization at low \acp{SNR}, even for higher-order constellations.
    \item We show that our proposed optimal detector also performs well in a realistic indoor \ac{THz} channel with multipath propagation and \ac{ISI}.
	\item We optimize the transmit symbol constellation for 1-bit and 2-bit quantization with respect to the \ac{SER}.
    The minimum achievable {\ac{SER}} can be decreased significantly by arranging the constellation symbols in a special way, making use of the nonlinear quantization operation.
\end{itemize}

This paper is organized as follows.
In Section~\ref{sec:sys_mod}, we introduce and motivate the system model.
In Section~\ref{sec:ADC_power}, the \ac{ADC} power consumption model is presented, and we specify \ac{ADC} parametrizations for a prescribed power consumption.
In Section~\ref{sec:symb_det}, we compute the statistics of the considered detection variable and present corresponding detectors, including the optimal \ac{ML} detector.
In Section~\ref{sec:num_results}, we present numerical results for the \ac{SER} of the presented detectors.
In Section~\ref{sec:const_opt}, we optimize the transmit constellation with respect to the \ac{SER} for 1- and 2-bit quantization.
Some conclusions are drawn in Section~\ref{sec:concl}.

\textit{Notation:}
$a$, $\lvec{a}$, and $\lmat{A}$ represent a scalar, a column vector, and a matrix, respectively.
$a_i$ is the $i^\text{th}$ element of the vector $\lvec{a}$.
$(\cdot)^*$ and $(\cdot)^T$ denote the complex conjugate and the transposition operation, respectively.
$\diag{\cdot}$ is a diagonal matrix with the elements in brackets on the main diagonal.
The Hadamard product is given by $\odot$.
$|\mathcal{A}|$ stands for the cardinality of the set $\mathcal{A}$.
$\est{\cdot}$ is the expectation operator and $\lfloor\cdot\rfloor$ represents the floor function.
The set of natural numbers including zero and the set of real numbers are denoted by $\Nset_0$ and $\Rset$, respectively.
$\lvec{0}_N$, $\lvec{1}_N$, and $\lmat{I}_N$ represent an all-zero column vector of length $N$, an all-ones column vector of length $N$, and an $N\times N$ identity matrix, respectively.
$\mathcal{R}\{\cdot\}$ and $\mathcal{I}\{\cdot\}$ return the real and imaginary part of the argument, respectively.
$\mathcal{N}(\gvec{\mu},\gmat{\Sigma})$ and $\mathcal{CN}(\gvec{\mu},\gmat{\Sigma})$ denote a real and complex multivariate Gaussian distribution with mean vector $\gvec{\mu}$ and covariance matrix $\gmat{\Sigma}$, respectively.
$Q(x)=\frac{1}{\sqrt{2\pi}}\int_x^\infty \er^{-\frac{t^2}{2}}dt$ is the Q-function.

\section{System Model}\label{sec:sys_mod}
\begin{figure}[t]
    \centerline{\includegraphics[width=\plotwidth\textwidth]{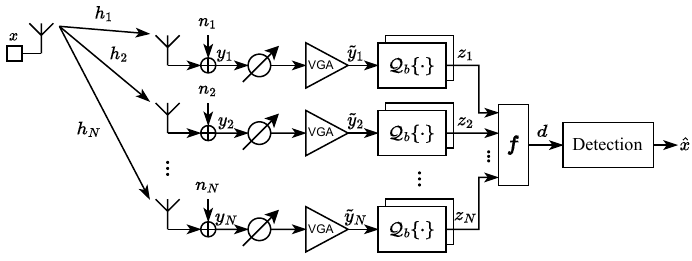}}
    \caption{\ac{SIMO} \ac{LoS} \ac{THz} channel with analog phase shifters, \aclp{VGA}, uniform quantizers, linear filter, and detector at the receiver side.}
    \label{fig:sys_mod}
\end{figure}

\begin{figure}[t]
    \centerline{\includegraphics[width=\plotwidth\textwidth]{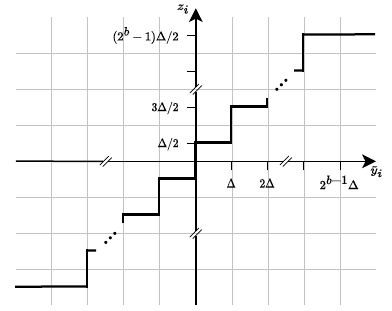}}
    \caption{Law $\mathcal{Q}_b\{\cdot\}$ of $b$-bit uniform midrise quantizer with step size $\Delta$.}
    \label{fig:uniform_b-bit_quantizer}
\end{figure}

We consider a \ac{LoS} \ac{THz} channel with uniform quantization and oversampling at the receiver side.
Oversampling is applied in the spatial domain, i.e., we consider a \ac{SIMO} system with multiple receive antennas.
The corresponding system model is illustrated in Fig.~\ref{fig:sys_mod}.
Here, the input symbol $\cmplx{x}\in\mathcal{X}$ is chosen from a complex finite alphabet $\mathcal{X}$ with cardinality $|\mathcal{X}|=M$ and power $\sigma_x^2=\est[\cmplx{x}]{|\cmplx{x}|^2}$.\footnote{We use the superscript $\cmplx{(\cdot)}$ to denote complex-valued variables. When dealing with real-valued quadrature components later on, we omit this superscript for better readability.}
More specifically, an $M$-ary square \ac{QAM} constellation with equiprobable symbols is considered.
For the following developments, we assume a frequency-flat single-path \ac{LoS} \ac{THz} channel which is a valid assumption since \ac{NLoS} paths are heavily attenuated due to the use of directional antennas and the additional reflection and scattering loss in the \ac{THz} band~\cite{Sarieddeen2021}.
Furthermore, we consider transmission windows which are not heavily affected by molecular absorption which renders them practically frequency flat.
However, these assumptions are discarded in Subsection~\ref{subsec:THz} where we study a realistic indoor \ac{THz} channel, modeled based on ray tracing.
The \ac{SIMO} channel with $N$ receive antennas can be described as a vector $\lvecg{h}=[\cmplx{h}_1\;\cdots\;\cmplx{h}_N]^T$  with channel coefficients~\cite{Jornet2011}
\begin{align}
    \cmplx{h}_i = \frac{c}{4\pi fd_i}\cdot\er^{-\frac{1}{2}\kappa(f)d_i}\cdot\er^{-\imag2\pi f\frac{d_i}{c}},
    \label{eq:channel_coeff}
\end{align}
where $c$ is the speed of light, $f$ is the operating frequency, $d_i$ denotes the distance between transmitter and receive antenna $i$, and $\kappa(f)$ stands for the frequency-dependent molecular absorption coefficient.
Thus, the first term in~\eqref{eq:channel_coeff} accounts for the spreading loss, the second term stands for the molecular absorption loss, and the last term indicates the phase shift the wave experiences when traveling over a distance $d_i$.
At the receiver, the signal is corrupted by \ac{AWGN} which yields the receive vector
\begin{align}
    \lvecg{y} = \lvecg{h}\cmplx{x} + \lvecg{n},
    \label{eq:rx_vec}
\end{align}
with $\lvecg{n}\sim\mathcal{CN}(\lvec{0}_N,\sigma_n^2\lmat{I}_N)$.
We define the \ac{SNR} as the ratio of the average received symbol power and the noise power, $\text{SNR}=\frac{||\lvecg{h}||_2^2\sigma_x^2}{N\sigma_n^2}$.
Before quantization, the noisy receive samples $\cmplx{y}_i$ are fed into analog phase shifters and \acp{VGA} which compensate for the channel's phase shift and attenuation.
For this, perfect knowledge of the channel coefficients $\cmplx{h}_i$ is assumed at the receiver.
The analog signal processing yields the vector
\begin{align}
    \lvecg{\tilde{y}} = {\lvecgconj{\tilde{h}}} \odot \lvecg{y} = \cmplx{x}\lvec{1}_N + \lvecg{\tilde{n}},
    \label{eq:rx_vec_asp}
\end{align}
with $\lvecg{\tilde{h}}=\left[\frac{\cmplx{h}_1}{|\cmplx{h}_1|^2}\cdots\frac{\cmplx{h}_N}{|\cmplx{h}_N|^2}\right]^T$ and $\lvecg{\tilde{n}}\sim\mathcal{CN}\Big(\lvec{0}_N,\text{diag}\Big\{\frac{\sigma_n^2}{|\cmplx{h}_1|^2},$ $\dots,\frac{\sigma_n^2}{|\cmplx{h}_N|^2}\Big\}\Big)$.
For co-located receive antennas, the different path gains are approximately the same, i.e., $|\cmplx{h}_1|^2\approx\dots\approx|\cmplx{h}_N|^2\eqcolon|\cmplx{h}|^2$.
Hence, the noise in different branches is \ac{iid}, $\lvecg{\tilde{n}}\sim\mathcal{CN}\left(\lvec{0}_N,\tilde{\sigma}_n^2\lmat{I}_N\right)$ with $\tilde{\sigma}_n^2=\frac{\sigma_n^2}{|\cmplx{h}|^2}$.
This yields $N$ independent \ac{AWGN} channels with the same noise variance.
Then, the vector $\lvecg{\tilde{y}}$ is quantized element-wise by a $b$-bit quantizer with law $\cmplx{\mathcal{Q}}_b\{\cmplx{\tilde{y}}_i\}=\mathcal{Q}_b\{\mathcal{R}\{\cmplx{\tilde{y}}_i\}\}+\imag\mathcal{Q}_b\{\mathcal{I}\{\cmplx{\tilde{y}}_i\}\}$, i.e., real and imaginary part are quantized separately.
The quantizer is chosen as a uniform midrise quantization law with step size $\Delta$,
\begin{align}
    \mathcal{Q}_b\{\tilde{y}_i\} = \begin{cases}
    \text{sign}(\tilde{y}_i)\cdot\left(\left\lfloor\frac{|\tilde{y}_i|}{\Delta}\right\rfloor\Delta+\frac{\Delta}{2}\right) &\text{for $|\tilde{y}_i|<2^{b-1}\Delta$}\\[10pt]
    \text{sign}(\tilde{y}_i)\cdot\frac{(2^b-1)\Delta}{2} &\text{otherwise}
    \end{cases},
    \label{eq:quantizer}
\end{align}
as illustrated in Fig.~\ref{fig:uniform_b-bit_quantizer}.
The resulting vector $\lvecg{z}$ is processed by a linear filter $\lvecg{f}\in\Rset^N$ which returns the detection variable
\begin{align}
    \cmplx{d} = {\lvecg{f}}^T\lvecg{z} = {\lvecg{f}}^T\cmplx{\mathcal{Q}}_b\{\lvecg{\tilde{y}}\} = {\lvecg{f}}^T\cmplx{\mathcal{Q}}_b\{\cmplx{x}\lvec{1}_N + \lvecg{\tilde{n}}\}.
    \label{eq:d_sys_mod}
\end{align}
We aim at analyzing the performance of the system for a specific detection filter $\lvecg{f} = \frac{1}{N}\lvec{1}_N$, constituting a simple averaging filter.
Finally, the detector returns the estimated input symbol $\cmplx{\hat{x}}$.

Note that the spatial-wideband effect~\cite{Wang2018} is not relevant in our scenario since the sampling time of each \ac{ADC} can be adjusted according to the path delay of the respective antenna.
Estimating these delays as well as the channel coefficients in~\eqref{eq:channel_coeff} based on quantized observations is beyond the scope of this paper.
However, there exist already approaches for such estimation in millimeter-wave channels under low-resolution quantization, cf. e.g.,~\cite{Kim2021,Vlachos2023,Doshi2023,Mewes2023}, based on compressive sensing techniques, the \ac{EM} algorithm, deep generative networks, and a combined {\ac{ML}} and least-squares approach, respectively.
It is expected that corresponding channel estimation approaches can be also designed for the \ac{THz} channels considered in this work.

Furthermore, our system model is also applicable to hybrid receivers which are often discussed in the literature.
Here, the signal after analog combining in a receiver subarray
would be fed into the \ac{VGA} before quantization. It should be noted that especially in the case of higher-order modulation, the availability of a sufficient number of quantized observations is crucial, i.e., analog combining should provide a sufficient number of output streams.

\section{ADC Power Consumption}\label{sec:ADC_power}
In this section, we introduce the adopted model for the power $P_\text{ADC}$ consumed in the \ac{ADC} based on the models in~\cite{Mezghani2010,Krone2011,Neuhaus2021,Castaneda2021} which can be unified as
\begin{align}
    P_\text{ADC} = \gamma\cdot N\cdot2^{\zeta q}\cdot f_s^\nu,
    \label{eq:P_ADC}
\end{align}
where $\gamma$ denotes a constant which depends on the utilized \ac{ADC} technology.
Here, one common choice for $\gamma$ is the Walden figure of merit~\cite{Walden1999} or the Schreier figure of merit~\cite{Schreier2005}.
Further information on the figures of merit and the different {\ac{ADC}} technologies such as flash, pipelined, \ac{SAR}, or sigma-delta {\acp{ADC}} can be found in~\cite{Murmann2015}.
Hence, the power consumption model presented in~{\eqref{eq:P_ADC}} is universally valid for different {\ac{ADC}} architectures.
Furthermore, $N$ stands for the oversampling factor as introduced in the previous section, $q$ denotes either the total number of bits $b$ used in quantization or the \ac{ENOB} of a given \ac{ADC}, the parameter $\zeta\in\{1,2\}$ depends on the quantization resolution, $f_s$ is the sampling frequency, and the parameter $\nu\in\{1,2\}$ is related to the sampling frequency.
For lower-resolution \acp{ADC}, $\zeta=1$ is usually chosen, whereas $\zeta=2$ is used for moderate-to-higher-resolution \acp{ADC}, and $\nu=1$ is utilized for small sampling frequencies while $\nu=2$ for larger sampling frequencies.
According to~\cite{Murmann2015} and~\cite{Murmann2022}, a doubling of the power consumption when increasing the quantization resolution by 1, i.e., $\zeta=1$, occurs for \acp{ADC} with a \ac{SNDR} of $\text{SNDR}<\SI{50}{dB}$ which is equivalent to $\text{ENOB}<\SI{8}{bit}$.
Larger sampling frequencies, i.e., $\nu=2$, correspond to the range $f_s>\SI{100}{MHz}$~\cite{Murmann2015,Murmann2022}.

In this work, we analyze the performance of low-resolution \acp{ADC} which are primarily of interest for very high sampling frequencies.
Hence, we consider the parametrization $\zeta=1$ and $\nu=2$.
Furthermore, we use $q=b$ which facilitates the power consumption computation of the $b$-bit quantizer introduced in the last section.\footnote{\ac{ENOB} depends on the actual hardware realization of the \ac{ADC}. Since we do not consider any specific \ac{ADC} realization, we use $q=b$.}
As a consequence of this parametrization, increasing the quantization resolution by 1 bit has the same effect as doubling the oversampling factor $N$.
Therefore, for example, 1-bit quantization ($b=1$) with an oversampling factor of $N=64$ results in the same power consumption as using a 2-bit \ac{ADC} with $N=32$ or a 3-bit \ac{ADC} with $N=16$, respectively.
These three cases will be analyzed in Section~\ref{sec:num_results} in more detail.
It should be noted that the linear scaling of the power consumption with respect to the oversampling factor does not contradict the quadratic increase with respect to the sampling frequency $f_s$ according to~(\ref{eq:P_ADC}) with $\nu=2$.
Here, the linear scaling with respect to $N$ can be motivated by time-interleaved \ac{ADC} architectures or by spatial oversampling.

\section{Optimal Symbol Detection}\label{sec:symb_det}
The optimal \ac{ML} detector maximizes the \ac{PMF} of the detection variable $\cmplx{d}$ given the input symbol $\cmplx{x}$, i.e., $p_{\cmplx{d}|\cmplx{x}}(\cmplx{d}|\cmplx{x})$.
The resulting \ac{ML} estimate of the transmitted symbol is
\begin{align}
    \cmplx{\hat{x}} = \argmax_{\cmplx{x}\in\mathcal{X}}p_{\cmplx{d}|\cmplx{x}}(\cmplx{d}|\cmplx{x}).
    \label{eq:ML_det_complex}
\end{align}
In order to derive the likelihood function $p_{\cmplx{d}|\cmplx{x}}(\cmplx{d}|\cmplx{x})$, the conditional probability distributions of the unquantized and quantized received signals have to be determined first.
For this, we make the assumption of co-located receive antennas which yields approximately \ac{iid} noise samples.
Furthermore, we consider the real and the imaginary part separately which is justified since both components are independent and have the same noise variance according to~\eqref{eq:rx_vec_asp}.
In the following, we express the fact that we consider only one quadrature component by omitting the superscript $\cmplx{(\cdot)}$, e.g., $\iq{x}$ and $\iq{d}$ denote either the in-phase or the quadrature component of the input symbol $\cmplx{x}$ and the detection variable $\cmplx{d}$, respectively.
The \ac{ML} estimate of the respective quadrature components is, hence, given by
\begin{align}
    \hat{\iq{x}} = \argmax_{\iq{x}\in\mathcal{X}'}p_{\iq{d}|\iq{x}}(\iq{d}|\iq{x}),
    \label{eq:ML_det}
\end{align}
with the $M'$-ary quadrature component constellation $\mathcal{X}'=\mathcal{R}\{\mathcal{X}\}=\mathcal{I}\{\mathcal{X}\}$ of the original $M$-ary square \ac{QAM} constellation $\mathcal{X}$ where $M'\coloneq\sqrt{M}$.

\subsection{Probability Distributions of Received Signals and Detection Variable}\label{subsec:prob_distr}
For a fixed input symbol $\iq{x}$, the unquantized vector $\tilde{\lvec{y}}$ follows the same probability distribution as the noise vector except for a shift of the mean by $\iq{x}$, i.e., $\tilde{\lvec{y}}\sim\mathcal{N}(\iq{x}\lvec{1}_N,\frac{\tilde{\sigma}_n^2}{2}\lmat{I}_N)$ which implies that the unquantized samples are \ac{iid} Gaussian distributed with mean $\iq{x}$ and variance $\frac{\tilde{\sigma}_n^2}{2}$.
We denote the random variable corresponding to one received sample $\tilde{\iq{y}}_n$ as $\tilde{\iq{y}}$.
Due to the independent unquantized observations and the element-wise quantization, the quantized vector $\lvec{z}$ also consists of independent entries if conditioned on $\iq{x}$.
Therefore, the corresponding likelihood function can be written as
\begin{align}
    p_{\lvec{z}|\iq{x}}(\lvec{z}|\iq{x}) = \prod_{n=1}^{N} p_{\iq{z}|\iq{x}}(\iq{z}_n|\iq{x}),
    \label{eq:z_likelihood}
\end{align}
where $p_{\iq{z}|\iq{x}}(\iq{z}_n|\iq{x})$ is the \ac{PMF} of the $n^\text{th}$ entry of $\lvec{z}$ given $\iq{x}$, $n\in\{1,\dots,N\}$.
Here, the quantized samples $\iq{z}_n$ are realizations of the discrete random variable $\iq{z}$ which follows a multinoulli distribution with $K=2^b$ different values $\iq{z}(k)\coloneqq\left(k-\frac{K+1}{2}\right)\Delta$, $k\in\{1,\dots,K\}$.
Due to the normally distributed elements in the unquantized vector, the conditional probabilities of the possible values of the random variable $\iq{z}$ can be computed as
\begin{align}
\begin{split}
    P_k(\iq{x}) &\coloneqq p_{\iq{z}|\iq{x}}(\iq{z}(k)|\iq{x}) = p_{\tilde{\iq{y}}|\iq{x}}(\tilde{\iq{y}}\in\mathcal{D}_{\tilde{\iq{y}}}(\iq{z}(k))|\iq{x})\\[5pt]
    &= Q\left(\frac{\tau_\text{low}(\iq{z}(k))-\iq{x}}{\tilde{\sigma}_n/\sqrt{2}}\right)-Q\left(\frac{\tau_\text{up}(\iq{z}(k))-\iq{x}}{\tilde{\sigma}_n/\sqrt{2}}\right),
    \label{eq:z_n_probability}
\end{split}
\end{align}
where the quantizer decision region and corresponding thresholds are given by
\begin{align}
    \mathcal{D}_{\tilde{\iq{y}}}(\iq{z}(k)) &= \{\tilde{\iq{y}}\in\Rset\,|\,\tau_\text{low}(\iq{z}(k))\leq \tilde{\iq{y}}<\tau_\text{up}(\iq{z}(k))\},\label{eq:decision_region_quantizer}\\[5pt]
    \tau_\text{low}(\iq{z}(k)) &= \begin{cases}
    \iq{z}(k)-\frac{\Delta}{2} &\text{for $\iq{z}(k)>-\frac{(K-1)\Delta}{2}$}\label{eq:decision_threshold_low_quantizer}\\
    -\infty &\text{otherwise}
    \end{cases},
\end{align}
and
\begin{align}
	\tau_\text{up}(\iq{z}(k)) &= \begin{cases}
    \iq{z}(k)+\frac{\Delta}{2} &\text{for $\iq{z}(k)<\frac{(K-1)\Delta}{2}$}\\
    +\infty &\text{otherwise}
    \end{cases}.
    \label{eq:decision_threshold_up_quantizer}
\end{align}
Now, the mean and the variance of $\iq{z}$ given $\iq{x}$ can be calculated as
\begin{align}
\begin{split}
    \mu_\iq{z}(\iq{x}) &= \est[\iq{z}|\iq{x}]{\iq{z}|\iq{x}}\\
    &= \Delta\cdot\left(\sum_{k=1}^{K-1}Q\left(\frac{\tau_\text{up}(\iq{z}(k))-\iq{x}}{\tilde{\sigma}_n/\sqrt{2}}\right)-\frac{K-1}{2}\right),
\end{split}\label{eq:z_n_mean}\\
\begin{split}
    \sigma_\iq{z}^2(\iq{x}) &= \est[\iq{z}|\iq{x}]{(\iq{z}-\mu_\iq{z}(\iq{x}))^2|\iq{x}}\\
    &= \Delta^2\cdot\Bigg(\sum_{k=1}^{K-1}(2k-K)\cdot Q\left(\frac{\tau_\text{up}(\iq{z}(k))-\iq{x}}{\tilde{\sigma}_n/\sqrt{2}}\right)\\
    &\phantom{=\Delta^2\cdot\Bigg(}+\left(\frac{K-1}{2}\right)^2\Bigg)-\mu_\iq{z}^2(\iq{x}),
\end{split}\label{eq:z_n_variance}
\end{align}
where we have used the fact that $\tau_\text{low}(\iq{z}(k+1))=\tau_\text{up}(\iq{z}(k))$ for $k\in\{1,\dots,K-1\}$.

\begin{figure}[t]
    \centering
    \subfloat[$b=1$, $\Delta=2$, $N=64$.]{\includegraphics[width=\plotwidth\textwidth]{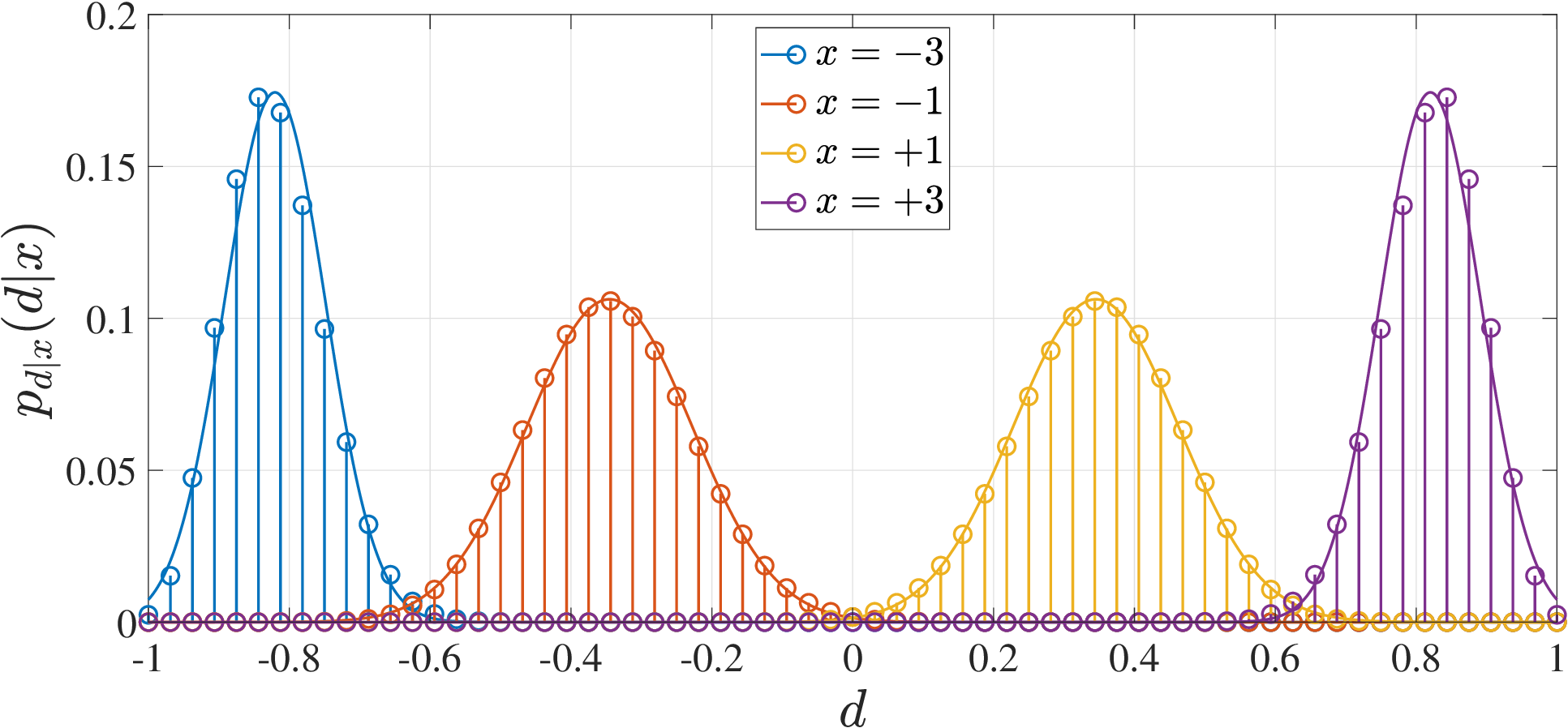}
    \label{fig:d_distr_1bit_4ASK}}\\
    \subfloat[$b=3$, $\Delta=1$, $N=16$.]{\includegraphics[width=\plotwidth\textwidth]{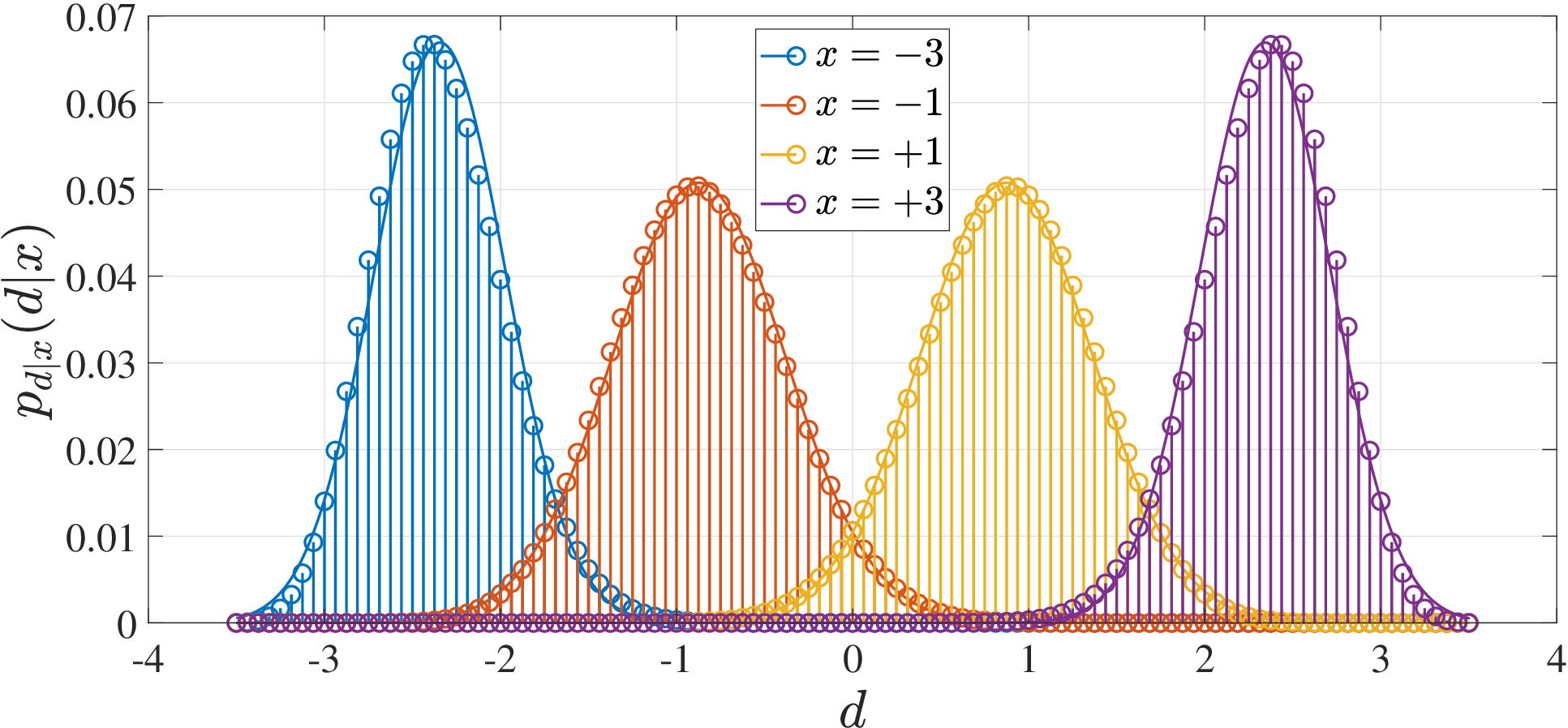}
    \label{fig:d_distr_3bit_4ASK}}\\
    \caption{{\acs{PMF}} $p_{\iq{d}|\iq{x}}(\iq{d}|\iq{x})$~({\ref{eq:d_pmf}}) and (scaled) continuous approximation $f_{\iq{d}|\iq{x},\text{CLT}}(\iq{d}|\iq{x})$~({\ref{eq:d_pdf_clt}}) for $\mathcal{X}'=\{\pm1,\pm3\}$ at $\text{SNR}=0\,$dB.}
    \label{fig:d_distr_4ASK}
\end{figure}

As mentioned above, the random variable $\iq{z}$ follows a multinoulli distribution with $K$ possible values and is observed $N$ times.
We denote the number of observations of the $k^\text{th}$ value $\iq{z}(k)$ as $\kappa_k\in\{0,\dots,N\}$ with $\sum_{k=1}^K\kappa_k=N$.
Hence, the collection of the numbers of observations is multinomially distributed,
\begin{align}
    p_{\gvec{\kappa}|\iq{x}}(\gvec{\kappa}|\iq{x}) = \binom{N}{\kappa_1,\dots,\kappa_K}\prod_{k=1}^KP_k^{\kappa_k}(\iq{x}),
    \label{eq:kappa_pmf}
\end{align}
with the vector of numbers of observations $\gvec{\kappa}=[\kappa_1\cdots\kappa_K]$ and the multinomial coefficient $\binom{N}{\kappa_1,\dots,\kappa_K}=\frac{N!}{\kappa_1!\dots\kappa_K!}$.
Since the detection variable $\iq{d}$ is the average of all $N$ observations, it can be expressed in terms of the numbers of observations $\kappa_k$,
\begin{align}
    \iq{d}=\iq{d}(\gvec{\kappa})=\frac{1}{N}\sum_{k=1}^K\kappa_k\iq{z}(k)=\frac{1}{N}\sum_{k=1}^K\kappa_k\left(k-\frac{K+1}{2}\right)\Delta.
    \label{eq:d_of_kappa}
\end{align}
Equation~(\ref{eq:d_of_kappa}) can be rearranged as
\begin{align}
    \sum_{k=1}^K\kappa_kk=N\left(\frac{d}{\Delta}+\frac{K+1}{2}\right).
    \label{eq:d_condition_kappa}
\end{align}
Therefore, all vectors $\gvec{\kappa}$ which satisfy~(\ref{eq:d_condition_kappa}) for fixed $\iq{d}$ yield the same value $\iq{d}$ of the detection variable.
Hence, the \ac{PMF} of the detection variable is given by a sum of multinomial distributions,
\begin{align}
    p_{\iq{d}|\iq{x}}(\iq{d}=\iq{d}(\gvec{\kappa})|\iq{x}) = \sum_{\gvec{\kappa}\in\mathcal{K}_\iq{d}}\binom{N}{\kappa_1,\dots,\kappa_K}\prod_{k=1}^KP_k^{\kappa_k}(\iq{x}),
    \label{eq:d_pmf}
\end{align}
with the set of vectors of numbers of observations corresponding to detection variable value $\iq{d}$,
\begin{align}
    \mathcal{K}_\iq{d}=\left\{\gvec{\kappa}\!\in\!\Nset_0^K\left|\sum_{k=1}^K\kappa_k\!=\!N\land\sum_{k=1}^K\kappa_kk\!=\!N\!\left(\!\frac{\iq{d}}{\Delta}\!+\!\frac{K+1}{2}\!\right)\!\right.\right\}.
    \label{eq:set_kappa_d}
\end{align}
Note that the detection variable can only take on discrete values from $\iq{z}(k=1)$ to $\iq{z}(k=K)$ with a spacing of $\frac{\Delta}{N}$, resulting in $N(K-1)+1$ possible different values.
The set of all possible vectors of numbers of observations $\mathcal{K}=\{\gvec{\kappa}\in\Nset_0^K|\sum_{k=1}^K\kappa_k=N\}$ consists of $|\mathcal{K}|=\binom{N+K-1}{K-1}$ different elements~\cite{Feller1968}.
If $b>1$ and $N>1$, there are more elements in $\mathcal{K}$ than possible detection variable values.
Hence, there is no one-to-one mapping between $\gvec{\kappa}$ and $\iq{d}$ for multi-bit quantization under oversampling which is exemplified in Table~\ref{tab:example} for $b=2$ and $N=2$.
\begin{table}[tb]
\caption{All possible vectors of numbers of observations and corresponding detection variable values for $b=2$ and $N=2$.}
\begin{center}
\begin{tabular}{|c|c|c|c|c|}
\hline
$\kappa_1$&$\kappa_2$&$\kappa_3$&$\kappa_4$&$\iq{d}$ \\
\hline
$2$&$0$&$0$&$0$&$-3/2\,\Delta$ \\
\hline
$1$&$1$&$0$&$0$&$-\Delta$ \\
\hline
$0$&$2$&$0$&$0$&$-1/2\,\Delta$ \\
\hline
$1$&$0$&$1$&$0$&$-1/2\,\Delta$ \\
\hline
$1$&$0$&$0$&$1$&$0$ \\
\hline
$0$&$1$&$1$&$0$&$0$ \\
\hline
$0$&$1$&$0$&$1$&$1/2\,\Delta$ \\
\hline
$0$&$0$&$2$&$0$&$1/2\,\Delta$ \\
\hline
$0$&$0$&$1$&$1$&$\Delta$ \\
\hline
$0$&$0$&$0$&$2$&$3/2\,\Delta$ \\
\hline
\end{tabular}
\label{tab:example}
\end{center}
\end{table}
This means that the receiver is not able to unambiguously determine the vector of numbers of observations $\gvec{\kappa}$ when inspecting the detection variable $\iq{d}$, contrary to 1-bit quantization where always a one-to-one mapping between $\gvec{\kappa}$ and $\iq{d}$ exists.
This has an influence on the complexity and the performance of the \ac{ML} detector based on $\iq{d}$ which will be discussed later on in more detail.
The mean and the variance of $\iq{d}$ can be computed using~(\ref{eq:z_n_mean}) and~(\ref{eq:z_n_variance}),
\begin{align}
    \mu_\iq{d}(\iq{x}) &= \est[\iq{d}|\iq{x}]{\iq{d}|\iq{x}} = \frac{1}{N}\sum_{n=1}^N\mu_\iq{z}(\iq{x}) = \mu_\iq{z}(\iq{x}),
    \label{eq:d_mean}\\
    \sigma_\iq{d}^2(\iq{x}) &= \est[\iq{d}|\iq{x}]{(\iq{d}\!-\!\mu_\iq{d}(\iq{x}))^2|\iq{x}} = \frac{1}{N^2}\sum_{n=1}^N\sigma_\iq{z}^2(\iq{x}) = \frac{1}{N}\sigma_\iq{z}^2(\iq{x}).
    \label{eq:d_variance}
\end{align}
Thus, a higher oversampling factor leads to a smaller variance of the detection variable.
With the above calculated mean and variance, via the \ac{CLT}, we can approximate the \ac{PMF} of $\iq{d}$ for large $N$ with a continuous Gaussian \ac{PDF}, $\iq{d}\sim\mathcal{N}(\mu_\iq{d}(\iq{x}),\sigma_\iq{d}^2(\iq{x}))$,
\begin{align}
    f_{\iq{d}|\iq{x},\text{CLT}}(\iq{d}|\iq{x}) = \frac{1}{\sqrt{2\pi}\sigma_\iq{d}(\iq{x})}\;\er^{-\frac{(\iq{d}-\mu_\iq{d}(\iq{x}))^2}{2\,\sigma_\iq{d}^2(\iq{x})}}.
    \label{eq:d_pdf_clt}
\end{align}
In Fig.~\ref{fig:d_distr_1bit_4ASK}, the \ac{PMF} of $\iq{d}$ as well as the continuous \ac{CLT} approximation which is scaled with $\frac{\Delta}{N}$ are illustrated for $b=1$, $\Delta=2$, $N=64$, and a 4-\ac{ASK} constellation, corresponding to 16-\ac{QAM} when considering both quadrature components.
It can be seen that the approximation and the true distribution match very well for the given oversampling factor, input constellation, and \ac{SNR}.
However, there are some limitations to the continuous approximation of the discrete system which will be pointed out later.
Furthermore, the variance of $\iq{d}$ for input symbols with different magnitudes is not equal, resulting in consequences for the \ac{ML} detector which will be discussed in the next subsection.
In Fig.~\ref{fig:d_distr_3bit_4ASK}, the \ac{PMF} of $\iq{d}$ is shown for $b=3$, $\Delta=1$, $N=16$, and otherwise the same parameters as in Fig.~\ref{fig:d_distr_1bit_4ASK}.
Here, the variance of $\iq{d}$ for different input symbols varies as well.
However, the relative difference in the variances is smaller than for 1-bit quantization. 

\subsection{Symbol Detectors}\label{subsec:symb_det}
\begin{figure}[t]
    \centerline{\includegraphics[width=\plotwidth\textwidth]{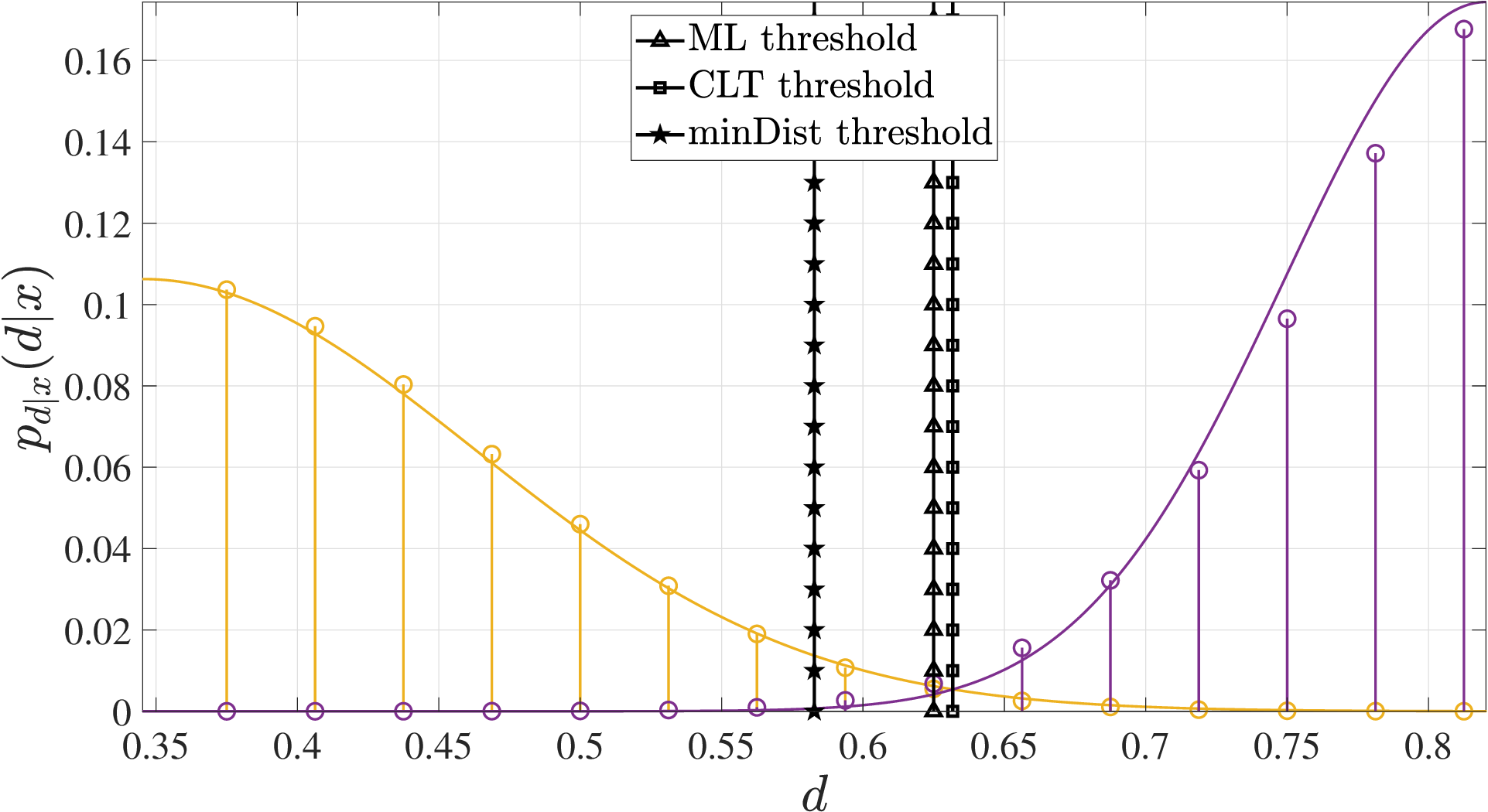}}
    \caption{Thresholds between the decision regions for $\iq{x}=+1$ and $\iq{x}=+3$ according to the detectors~(\ref{eq:ML_det_exact}),~(\ref{eq:ML_det_CLT}), and~(\ref{eq:subopt_det}) for $b=1$, $\Delta=2$, $N=64$, and $\mathcal{X}'=\{\pm1,\pm3\}$ at $\text{SNR}=0\,$dB. The colored discrete markers and continuous lines represent the \acs{PMF} $p_{\iq{d}|\iq{x}}(\iq{d}|\iq{x})$~(\ref{eq:d_pmf}) and (scaled) continuous approximation $f_{\iq{d}|\iq{x},\text{CLT}}(\iq{d}|\iq{x})$~(\ref{eq:d_pdf_clt}), respectively.}
    \label{fig:d_dec_thr}
\end{figure}

We can use the \ac{PMF} in~(\ref{eq:d_pmf}) to specify the \ac{ML} detector according to~(\ref{eq:ML_det}),
\begin{align}
    \hat{\iq{x}}_\text{ML} = \argmax_{\iq{x}\in\mathcal{X}'}\sum_{\gvec{\kappa}\in\mathcal{K}_d}\binom{N}{\kappa_1,\dots,\kappa_K}\prod_{k=1}^KP_k^{\kappa_k}(\iq{x}).
    \label{eq:ML_det_exact}
\end{align}
More details about the detector in~{\eqref{eq:ML_det_exact}} are discussed in the next subsection.
Moreover, we consider an alternative detector based on the \ac{CLT} approximation which is given by
\begin{align}
    \hat{\iq{x}}_\text{CLT} = \argmax_{\iq{x}\in\mathcal{X}'}\frac{1}{\sigma_\iq{d}(\iq{x})}\;\er^{-\frac{(\iq{d}-\mu_\iq{d}(\iq{x}))^2}{2\sigma_\iq{d}^2(\iq{x})}}.
    \label{eq:ML_det_CLT}
\end{align}
In our case, we cannot employ the classical \ac{ML} approach for an unquantized channel and minimize the Euclidean distance between $\iq{d}$ and $\mu_\iq{d}(\iq{x})$ in order to realize~(\ref{eq:ML_det_CLT}) because the variance $\sigma_\iq{d}^2(\iq{x})$ depends on $\iq{x}$.
Therefore, in general, the optimal decision thresholds do not lie exactly in the middle between the mean values $\mu_\iq{d}(\iq{x})$ of adjacent input symbols.
Nevertheless, we investigate the performance of such suboptimal approach as well with decision rule
\begin{align}
    \hat{\iq{x}}_\text{minDist} = \argmin_{\iq{x}\in\mathcal{X}'}\left(\iq{d}-\mu_\iq{d}(\iq{x})\right)^2.
    \label{eq:subopt_det}
\end{align}
The three detectors according to~(\ref{eq:ML_det_exact}),~(\ref{eq:ML_det_CLT}), and~(\ref{eq:subopt_det}) result in three different sets of decision thresholds for the scalar detection variable $d$.
In Fig.~\ref{fig:d_dec_thr}, the thresholds between decision regions for $\iq{x}=+1$ and $\iq{x}=+3$ are depicted exemplarily for 1-bit quantization.

The fully optimum \ac{ML} detector is directly based on the quantized vector $\lvec{z}$, i.e., no filter $\lvecg{f}$ is applied, and can be expressed via the likelihood function in~(\ref{eq:z_likelihood}), resulting in
\begin{align}
    \hat{\iq{x}}_\text{noFilt} = \argmax_{\iq{x}\in\mathcal{X}'}p_{\lvec{z}|\iq{x}}(\lvec{z}|\iq{x}) = \argmax_{\iq{x}\in\mathcal{X}'}\prod_{k=1}^KP_k^{\kappa_k}(\iq{x}).
    \label{eq:z_ML_det}
\end{align}
According to~(\ref{eq:z_ML_det}), the exact vector $\lvec{z}$ is not relevant for detection but the numbers of observations $\kappa_k$, $k\in\{1,\dots,K\}$, are a set of sufficient statistics which was also mentioned in~\cite{Dabeer2006} for 1-bit quantization.
Due to the one-to-one mapping between $\iq{d}$ and $\gvec{\kappa}$ in case of 1-bit quantization, which makes the averaging filtering an invertible operation, it can be easily shown via the data processing theorem~\cite{Cover2006} that the detector in~(\ref{eq:ML_det_exact}) is equivalent to the fully optimum detector in~(\ref{eq:z_ML_det}).
As a consequence, the averaging filter $\lvecg{f}$ is optimal for the considered system with $b=1$.
For $b>1$ and $N>1$, however, this is not the case anymore due to the many-to-one mapping between the numbers of observations and the detection variable, exemplified in Table~\ref{tab:example}.
Therefore, for multi-bit quantization with oversampling, the detector in~(\ref{eq:ML_det_exact}) performs worse than the optimal detector in~(\ref{eq:z_ML_det}).
However, the performance loss is negligible as will be observed later from the numerical results.

\subsection{Computational and Hardware Complexity}\label{subsec:complexity}
In this subsection, we discuss the computational complexity of the aforementioned detection schemes.
Furthermore, we consider different hardware implementation aspects and corresponding challenges.

All the detection schemes in~({\ref{eq:ML_det_exact}}),~({\ref{eq:ML_det_CLT}}), and~({\ref{eq:subopt_det}}) utilize the detection variable $d$ for their decision which is obtained through the averaging filter requiring the summation of $N$ quantized values.
For detecting a quadrature component of the transmitted symbol, the space of possible values for $d$ can be partitioned by $M'-1$ decision boundaries into $M'$ decision regions, each one indicating a different detected quadrature component symbol.
Under the assumption that the decision boundaries are given, the computational complexity per quadrature component of the aforementioned detectors is determined by the comparison of the detection variable $d$ to the $M'-1$ decision thresholds.
Hence, the detectors presented in~({\ref{eq:ML_det_exact}}),~({\ref{eq:ML_det_CLT}}), and~({\ref{eq:subopt_det}}) exhibit a low computational complexity.
Computing the decision thresholds comprises a higher complexity.
However, the corresponding computations can be done offline before detecting the symbols.
This will be discussed in more detail below.
The complexity per quadrature component of the fully optimum {\ac{ML}} detector without filtering according to~{\eqref{eq:z_ML_det}} is given by the $M'$-fold computation of the product of $K$ powers which is equivalent to the product of $N$ probability values since $\sum_{k=1}^K\kappa_k=N$.
Hence, the detection complexity is higher compared to that of the detectors presented in~({\ref{eq:ML_det_exact}}),~({\ref{eq:ML_det_CLT}}), and~({\ref{eq:subopt_det}}).

In the following, we discuss how to obtain the $M'-1$ decision boundaries $[b_1\;b_2\;\cdots\;b_{M'-1}]$ and the corresponding computational complexity of the detectors utilizing the scalar detection variable $d$ for their decision.
In order to deter-mine the decision thresholds for the {\ac{ML}} detector in~({\ref{eq:ML_det_exact}}), the {\ac{PMF}} in~{(\ref{eq:d_pmf})} needs to be computed for each quadrature component symbol $\iq{x}\in\mathcal{X}'$ which requires the summation of $|\mathcal{K}| = \sum_{\forall d}|\mathcal{K}_d|=\binom{N+K-1}{K-1}$ terms, each comprising a multinomial coefficient and a product of $K=2^b$ powers.
This yields a high computational complexity for large oversampling factors $N$ and quantization resolutions $b$.
Without loss of generality, an ordering of the constellation symbols $\iq{x}_i\in\mathcal{X'}$ can be assumed, i.e., $\iq{x}_1<\iq{x}_2<\ldots<\iq{x}_{M'}$.
The decision threshold $b_i$ can be found by comparing the {\acp{PMF}} for the two adjacent input symbols $\iq{x}_i$ and $\iq{x}_{i+1}$, and choosing the smallest argument value of $\iq{d}$ where the function value corresponding to the greater symbol $\iq{x}_{i+1}$ is higher than or equal to the function value corresponding to the smaller symbol $\iq{x}_i$.
For determining the decision boundaries of the detector in~({\ref{eq:ML_det_CLT}}), the computation of the mean~({\ref{eq:d_mean}}) and variance~({\ref{eq:d_variance}}) of the {\ac{PDF}} in~({\ref{eq:d_pdf_clt}}) is necessary for each quadrature component symbol which requires the summation of $K-1$ terms according to~({\ref{eq:z_n_mean}}) and~({\ref{eq:z_n_variance}}), respectively.
The decision boundary $b_i$ is then found by closed-form computation of the intersection point of the two Gaussian distributions corresponding to the adjacent input symbols $\iq{x}_i$ and $\iq{x}_{i+1}$.
For the suboptimal detector~({\ref{eq:subopt_det}}), the $i^\text{th}$ decision threshold is given by $b_i=\frac{\mu_\iq{d}(\iq{x}_i)+\mu_\iq{d}(\iq{x}_{i+1})}{2}$ with $\mu_\iq{d}(\iq{x})$ from~({\ref{eq:d_mean}}) which requires the summation of $K-1$ terms as mentioned already earlier.
We note that all the above mentioned decision boundaries depend on the {\ac{SNR}}.
Thus, in practice, the decision boundaries need to be computed offline for several relevant {\ac{SNR}} values and then stored in the receiver's memory enabling low-complexity detection with high performance.
After estimation of {\ac{SNR}} at the receiver side, appropriate decision boundaries are selected.

In contrast to the threshold detection described above, the metrics given in~({\ref{eq:ML_det_exact}}),~({\ref{eq:ML_det_CLT}}), and~({\ref{eq:subopt_det}}), respectively, can be also directly computed and compared to estimate the transmitted symbol.
However, this yields a higher computational complexity, especially for large $N$ and $b$.
For example, the computational complexity of directly evaluating the {\ac{ML}} metric according to~{(\ref{eq:ML_det_exact})} is determined by the summation of $|\mathcal{K}_d|$ terms, each comprising a multinomial coefficient and a product of $K$ powers.
The exact number $|\mathcal{K}_d|$ depends on the value of the detection variable $d$.
However, since $|\mathcal{K}| = \sum_{\forall d}|\mathcal{K}_d|=\binom{N+K-1}{K-1}$, we can conclude that the number $|\mathcal{K}_d|$ increases with increasing oversampling factor $N$ and quantization resolution $b$.
For large $N$ and $b$, this detector requires a high computational complexity.
One special case which exhibits a low computational complexity is 1-bit quantization.
Here, the {\ac{ML}} detector in~{(\ref{eq:ML_det_exact})} simplifies significantly because $|\mathcal{K}_d|=1\;\forall \iq{d}$ due to the one-to-one mapping between $\gvec{\kappa}$ and $\iq{d}$, and, therefore, the multinomial coefficient in~({\ref{eq:ML_det_exact}}) is irrelevant for the maximization with respect to $\iq{x}$.
Hence, only the product of $K=2$ powers needs to be computed which was already indicated in~\cite{Forsch2022} and~\cite{Nakashima2018}.

Next, we comment on the hardware complexity of the pre-sented approach.
Due to the spatial oversampling considered in this work, the number of \ac{RF} chains in-creases linearly with the oversampling factor $N$.
One simplifi-cation can be made in the case of 1-bit quantization.
Here,~the {\acp{VGA}} in the block diagram in Fig.~{\ref{fig:sys_mod}} can be removed since for detection only the signs of the received signals are relevant.
Finally, we note that another limitation of the presented approach comes from the assumption of approximately identical path gains for every receive branch.
Due to this assumption, the antenna array size is limited since the path gains for different antennas can vary significantly for very large antenna arrays even when the antennas are co-located.

\subsection{Symbol Error Rate Computation}\label{subsec:SER}
In order to compute the \ac{SER} analytically for a given detector utilizing the detection variable $d$, $M'-1$ decision boundaries $[b_1\;b_2\;\cdots\;b_{M'-1}]$ have to be determined first as described in the last subsection.
Then, we define for each possible quadrature input symbol $\iq{x}_i\in\mathcal{X'}$ a decision region $\mathcal{D}_i$ for the detection variable which depends on the utilized detector.
Given the transmitted symbol is $\iq{x}_1$, a detection error occurs if the detection variable $\iq{d}$ is equal to or above the boundary $b_1$.
Thus, the decision region for $\iq{x}_1$ is $\mathcal{D}_1=[\iq{z}(k=1),b_1)$.
For the last symbol $\iq{x}_{M'}$, a wrong symbol will be detected if $\iq{d}$ is smaller than $b_{M'-1}$, thus $\mathcal{D}_{M'}=[b_{M'-1},\iq{z}(k=K)]$.
The decision region for any inner symbol $\iq{x}_i$, $i\in\{2,\dots,M'-1\}$, is given by $\mathcal{D}_i=[b_{i-1},b_i)$.
Correspondingly, the \ac{SER} of one quadrature component for all detectors based on $\iq{d}$ can be expressed as
\begin{align}
    \text{SER}_\text{ASK} = \frac{1}{M'}\sum_{\iq{x}_i\in\mathcal{X}'}\sum_{\iq{d}\notin\mathcal{D}_i}p_{\iq{d}|\iq{x}}(\iq{d}|\iq{x}_i),
    \label{eq:SER_ASK}
\end{align}
using the likelihood function $p_{\iq{d}|\iq{x}}(\iq{d}|\iq{x})$ in~(\ref{eq:d_pmf}).

In order to compute the \ac{SER} analytically for the detector utilizing $\lvec{z}$, we can evaluate the likelihood function in~(\ref{eq:z_likelihood}) for all possible quantized vectors $\lvec{z}$ and all possible input symbols $\iq{x}$ and then find suitable decision sets for all $\iq{x}$.
However, there are $K^N$ possible quantized vectors which might be too many for an efficient \ac{SER} computation.
Instead, we can compute the conditional probabilities of all possible vectors of numbers of observations $\gvec{\kappa}\in\mathcal{K}$ for each input symbol $\iq{x}$, $p_{\gvec{\kappa}|\iq{x}}(\gvec{\kappa}|\iq{x})$, according to~(\ref{eq:kappa_pmf}).
Then, we find the most probable symbol $\iq{x}_i$ for each vector $\gvec{\kappa}$ based on the computed probabilities and include the considered vector $\gvec{\kappa}$ in the decision set of the corresponding symbol $\iq{x}_i$, $\mathcal{K}_i=\{\gvec{\kappa}\in\mathcal{K}\,|\,p_{\gvec{\kappa}|\iq{x}}(\gvec{\kappa}|\iq{x}_i)\geq p_{\gvec{\kappa}|\iq{x}}(\gvec{\kappa}|\iq{x}_j)\;\forall j\neq i\}$, where we do not allow a vector $\gvec{\kappa}$ to belong to multiple decision sets.
The resulting \ac{SER} of one quadrature component is given by
\begin{align}
    \text{SER}_\text{ASK,noFilt} = \frac{1}{M'}\sum_{\iq{x}_i\in\mathcal{X}'}\sum_{\gvec{\kappa}\notin\mathcal{K}_i}p_{\gvec{\kappa}|\iq{x}}(\gvec{\kappa}|\iq{x}_i).
    \label{eq:SER_ASKnoFilt}
\end{align}

Due to the orthogonality of in-phase and quadrature components, the \ac{SER} of the complex-valued input symbols $\cmplx{x}\in\mathcal{X}$ is given by
\begin{align}
    \text{SER}=1-(1-\text{SER}')^2=2\cdot\text{SER}'-\text{SER}'^2,
    \label{eq:SER}
\end{align}
where $\text{SER}'$ is the \ac{SER} of one quadrature component, obtained from~(\ref{eq:SER_ASK}) or~(\ref{eq:SER_ASKnoFilt}).

\section{Numerical Results}\label{sec:num_results}
\subsection{Frequency-Flat Single-Path LoS THz Channel}\label{subsec:AWGN}
In the following, \ac{SER} results for the frequency-flat single-path \ac{LoS} \ac{THz} channel with co-located receive antennas, as introduced in Section~\ref{sec:sys_mod}, are presented for the derived detectors and the three \ac{ADC} parametrizations mentioned in Section~\ref{sec:ADC_power}.
Here, the exact choice of the transmit power, the carrier frequency, the distance between transmitter and receiver, or the noise variance is not important since the performance finally depends on the \ac{SNR}.
Therefore, we do not specify these individual parameters here.
The following results are valid for any combination of the aforementioned parameters which yield a given \ac{SNR} value.
A realistic choice of parameters for \ac{THz} communications is made in the next subsection.
Moreover, we have computed the \ac{SER} of an unquantized system with the same oversampling factor $N$ as ultimate benchmark which is equivalent to a single-antenna unquantized reception with noise variance ${\tilde{\sigma}_n^2}/{N}$.
In this case, the presented suboptimal decision boundaries become optimal since the variance of the detection variable for different input symbols become equal and minimizing the Euclidean distance is optimal.
Besides, we have conducted Monte Carlo simulations with $10^8$ random symbol transmissions in each case in order to verify the analytical results.
Furthermore, for 2- and 3-bit quantization, the choice of the step size $\Delta$ is important.
We heuristically choose different step sizes for $b=2$ and $b=3$, depending on the transmit symbol constellation.
For 16-\ac{QAM} transmissions with $\mathcal{X}'=\{\pm1,\pm3\}$, we choose the step size to be $\Delta=2$ for $b=2$ and $\Delta=1$ for $b=3$ such that the constellation symbols are evenly spread over the quantization region.
For 64-\ac{QAM} transmissions with $\mathcal{X}'=\{\pm1,\pm3,\pm5,\pm7\}$, we choose $\Delta=4$ for $b=2$ and $\Delta=2$ for $b=3$ for the same reason.
This choice of the step size is in fact optimal if the number of quantization intervals is larger than or equal to the number of different constellation symbols, i.e., for $b=2$ under 16-{\ac{QAM}} transmission as well as for $b=3$ for both considered constellations.
For higher-order constellations, e.g., for $b=2$ under 64-{\ac{QAM}} transmission, this choice is not optimal anymore but still yields good results.
The optimal choice of the step size in this case is related to the constellation optimization for higher-order constellations which is discussed in detail in Section~{\ref{sec:const_opt}}.

\begin{figure}[t]
    \centering
    \subfloat[$b=1$, $N=64$.]{\includegraphics[width=\plotwidth\textwidth]{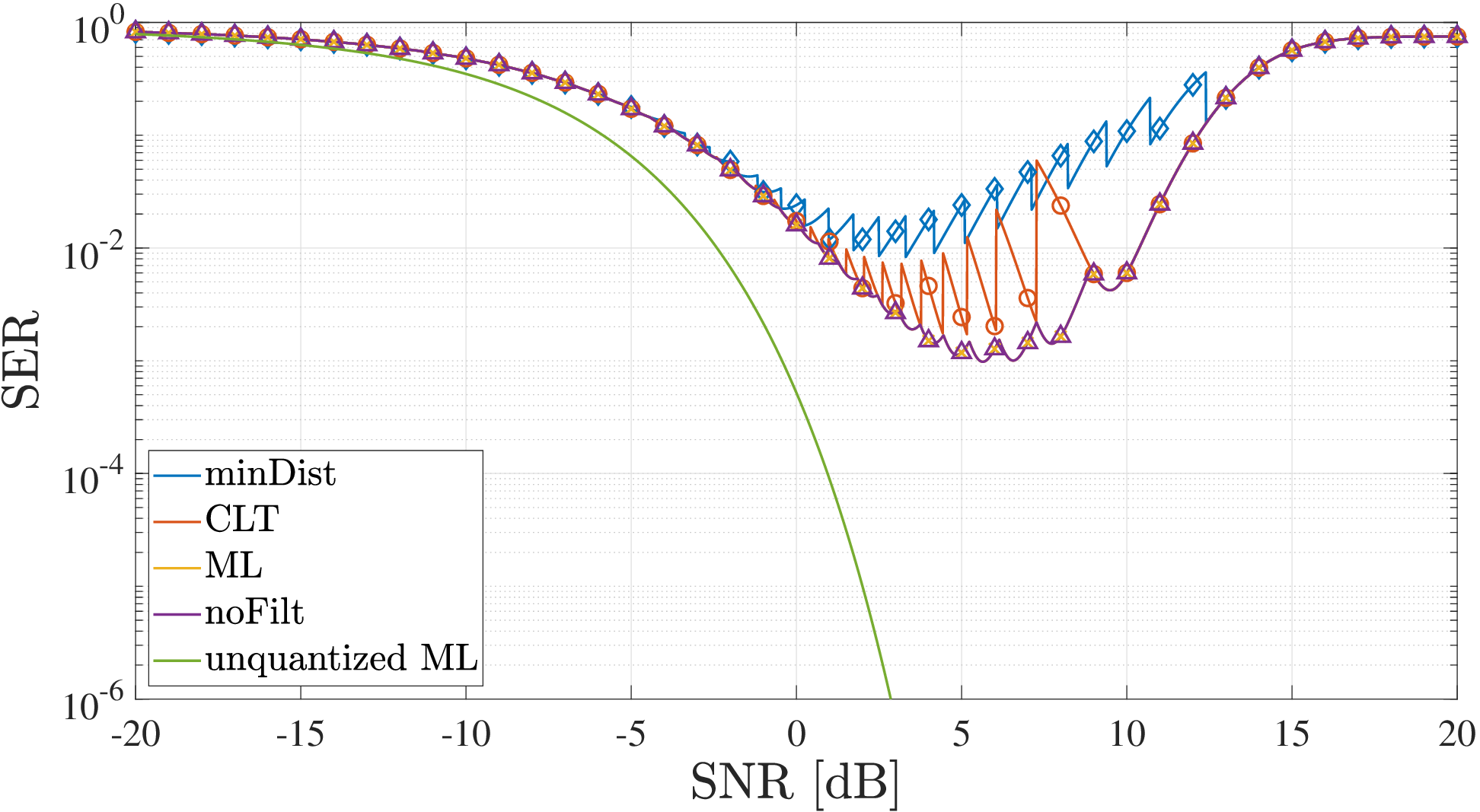}
        \label{fig:SER_1-bit_N64_4-ASK_detectors}}\\
    \subfloat[$b=2$, $N=32$, $\Delta=2$.]{\includegraphics[width=\plotwidth\textwidth]{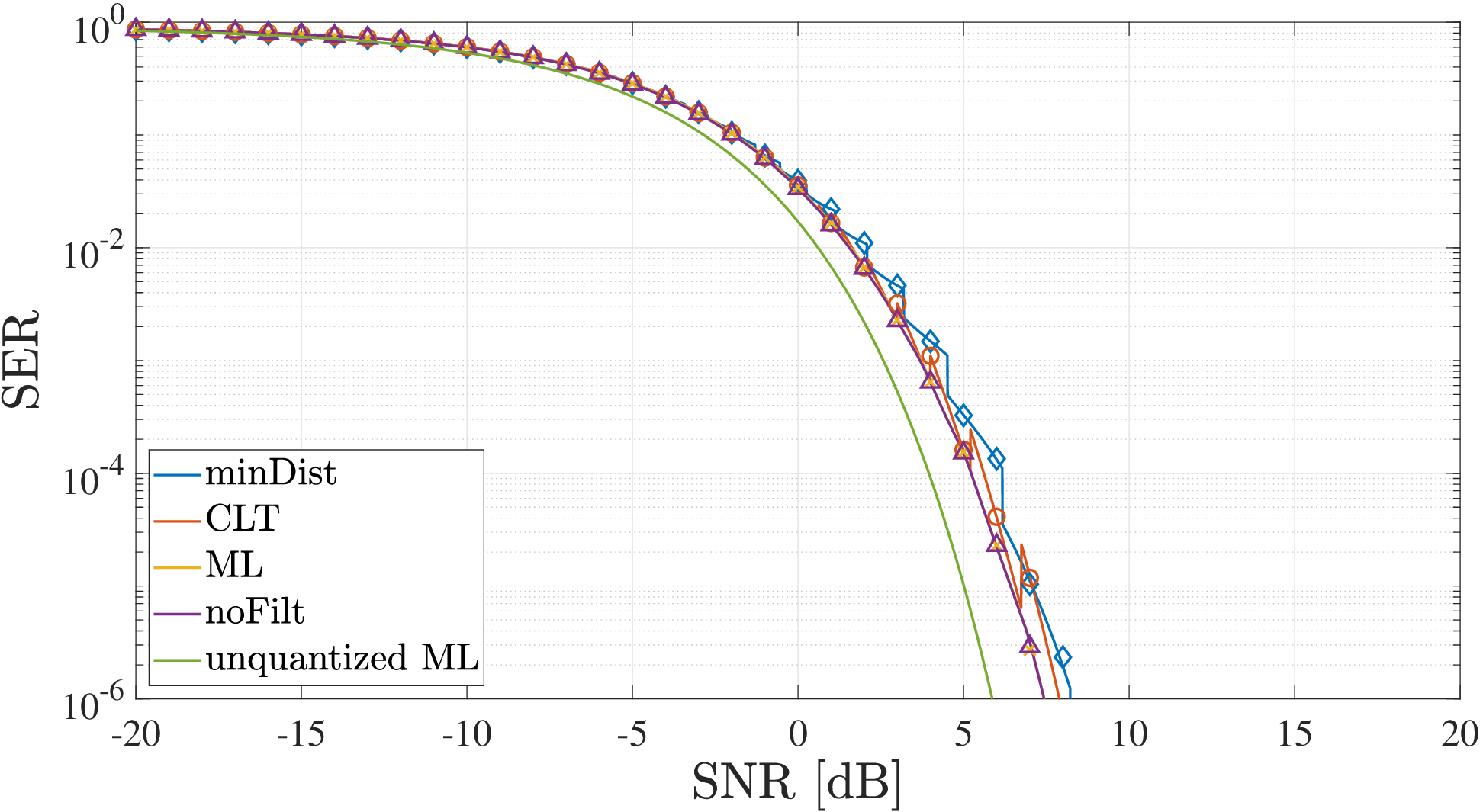}
        \label{fig:SER_2-bit_N32_4-ASK_detectors}}\\
    \subfloat[$b=3$, $N=16$, $\Delta=1$.]{\includegraphics[width=\plotwidth\textwidth]{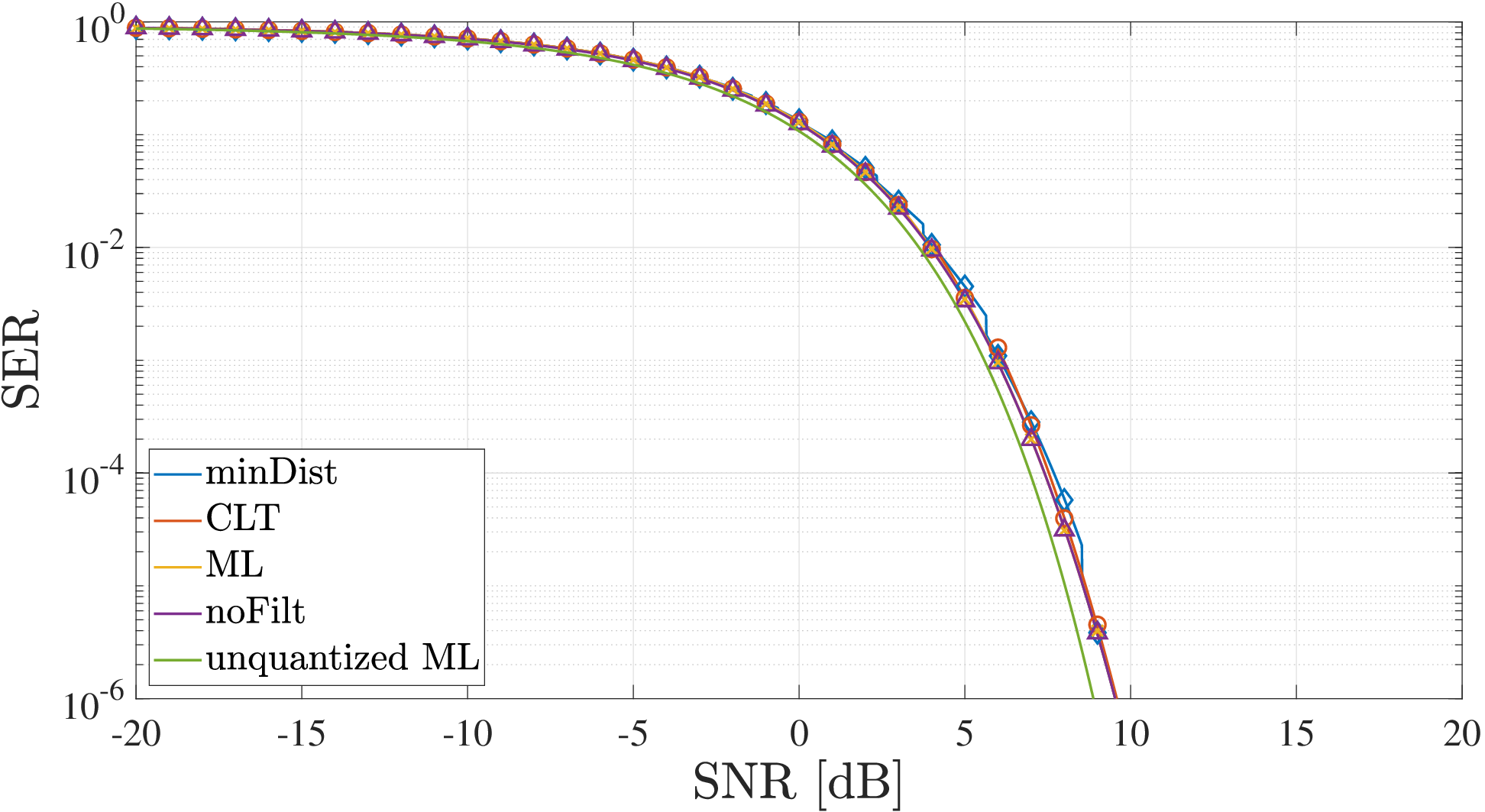}
        \label{fig:SER_3-bit_N16_4-ASK_detectors}}
    \caption{Analytical \acs{SER} (solid lines) and simulated \ac{SER} (markers) for the \acs{ML} detector~(\ref{eq:ML_det_exact}), CLT-based detector~(\ref{eq:ML_det_CLT}), minimum distance detector~(\ref{eq:subopt_det}), and fully optimum detector without filtering~(\ref{eq:z_ML_det}) under 16-\acs{QAM} transmission with $\mathcal{X}'=\{\pm1,\pm3\}$. Furthermore, the \ac{ML} detection performance for the unquantized case with the same oversampling factor is shown.}
    \label{fig:SER_4-ASK_detectors}
\end{figure}

At first, we consider a 16-\ac{QAM} constellation with $\mathcal{X}'=\{\pm1,\pm3\}$.
The corresponding \ac{SER} curves are illustrated in Fig.~\ref{fig:SER_4-ASK_detectors}.
For 1-bit quantization (Fig.~\ref{fig:SER_1-bit_N64_4-ASK_detectors}), we observe that the \ac{SER} curves are not smooth.
This behavior stems from the strong nonlinearity in the system and the discrete nature of the detection variable and can be only observed when the resolution with respect to the \ac{SNR}-axis, based on which the curves are drawn, is sufficiently high.
Using our analysis, the \ac{SER} results can be computed very fast for arbitrary \ac{SNR} values which is not feasible for exhaustive Monte Carlo simulations.
Furthermore, it can be observed that the \ac{SER} degrades for high \ac{SNR} and converges to 0.75 as expected due to the fact that at high \ac{SNR} the four symbols in each quadrant, e.g.,  $1\!+\!\imag$, $1\!+\!3\imag$, $3\!+\!\imag$ and $3\!+\!3\imag$ in the first quadrant, cannot be distinguished anymore since they are all mapped to the same quantization level due to the negligible influence of noise.
Additionally, one can observe from Fig.~\ref{fig:SER_1-bit_N64_4-ASK_detectors} that the decision rule based on the continuous \ac{CLT} approximation leads to a significant performance degradation compared to the optimal \ac{ML} detector.
The other suboptimal approach based on minimizing the Euclidean distance also leads to a significant performance degradation compared to the \ac{ML} detector.
Furthermore, it can be observed that the \ac{ML} detector based on the detection variable $\iq{d}$ in~(\ref{eq:ML_det_exact}) shows exactly the same performance as the fully optimum \ac{ML} detector based on the quantized vector $\lvec{z}$ in~(\ref{eq:z_ML_det}) which was already mentioned in Subsection~\ref{subsec:symb_det}.
Finally, the performance gap between 1-bit quantization and the unquantized case seems acceptable when considering the benefit of possible power savings in a real system.
Here, the loss in \ac{SNR} equals \SI{2.3}{dB}, \SI{3.2}{dB}, and \SI{6.0}{dB} at an error rate of $\text{SER}=10^{-1}$, $\text{SER}=10^{-2}$, and $\text{SER}=10^{-3}$, respectively.

\begin{figure}[tb]
    \centerline{\includegraphics[width=\plotwidth\textwidth]{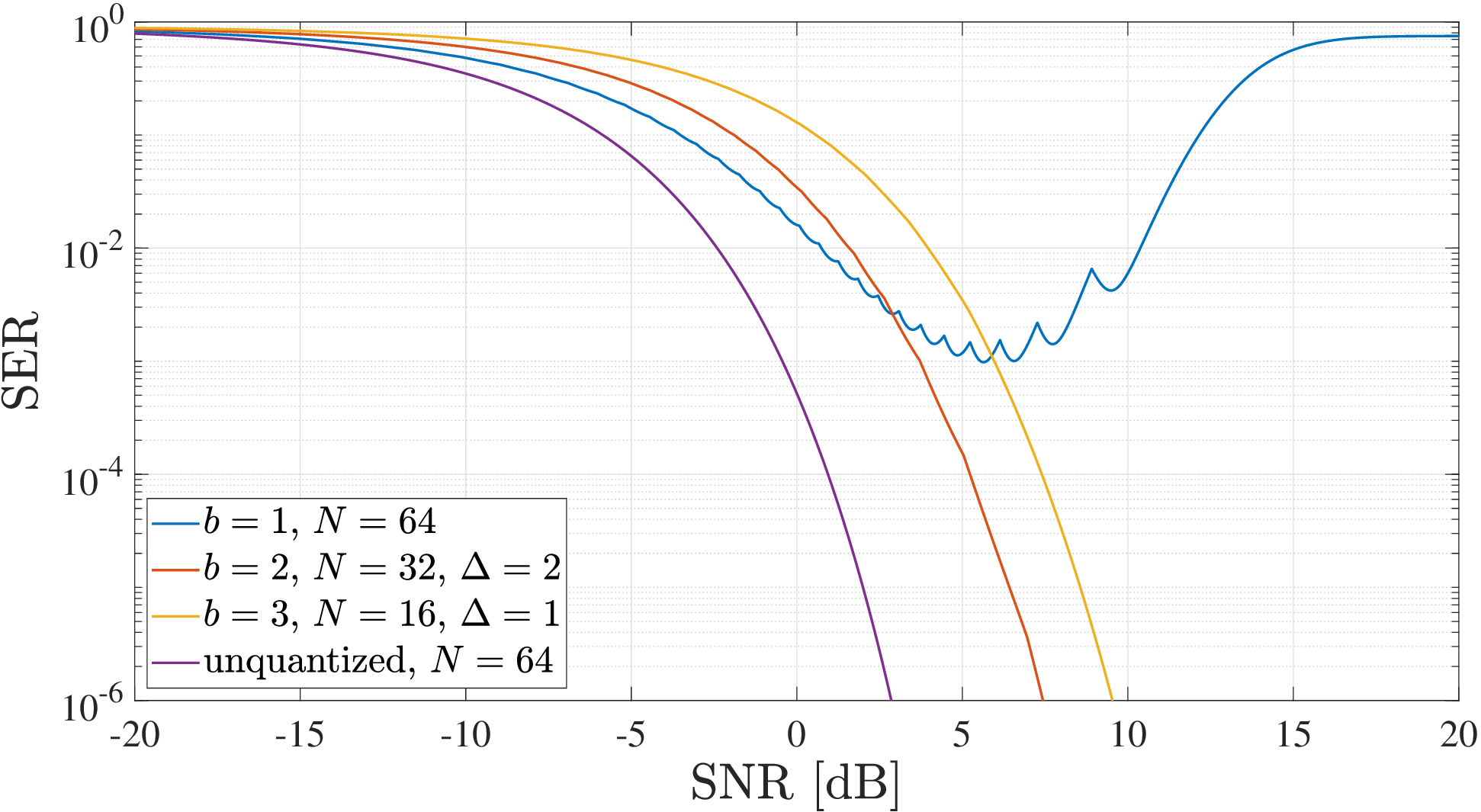}}
    \caption{Analytical {\acs{SER}} for the {\acs{ML}} detector~({\ref{eq:ML_det_exact}}) and different {\ac{ADC}} parametrizations under 16-{\ac{QAM}} transmission with $\mathcal{X}'=\{\pm1,\pm3\}$.}
    \label{fig:SER_1-3-bit_4-ASK_ML}
\end{figure}

For 2-bit quantization (Fig.~\ref{fig:SER_2-bit_N32_4-ASK_detectors}), a performance deviation between the optimal \ac{ML} detector based on $\iq{d}$ and both suboptimal detectors can be still observed even though the number of quantization levels is equal to the number of possible input symbols.
Only for 3-bit quantization (Fig.~\ref{fig:SER_3-bit_N16_4-ASK_detectors}), there is almost no noticeable difference between the three detectors.
Zooming in the curves for 2- and 3-bit quantization (not shown here), one observes a small performance loss of the \ac{ML} detector in~(\ref{eq:ML_det_exact}) compared to the fully optimum detector in~(\ref{eq:z_ML_det}).
This loss is related to the nonexistent one-to-one mapping between the vectors of numbers of observations $\gvec{\kappa}$ and the detection variable values $\iq{d}$ which was discussed in Subsection~\ref{subsec:prob_distr}.
However, the loss is negligibly small, thus, the combination of the quantized observations via an averaging filter does practically not degrade the system performance and yields near-optimum results.
Besides, it can be noticed for 2- and 3-bit quantization that the \ac{SER} goes to zero for increasing \ac{SNR} because the number of quantization levels is larger than or equal to the number of possible input symbols.
Nevertheless, there is still a performance loss compared to the unquantized case, even for 3-bit quantization.
Finally, it can be observed that the analytically obtained \ac{SER} results (solid lines) match exactly with the \ac{SER} results obtained by Monte Carlo simulations (markers) in all cases.
This confirms that our analysis is accurate.
In Fig.~\ref{fig:SER_1-3-bit_4-ASK_ML}, the performance of the different \ac{ADC} parametrizations is compared for the \ac{ML} detector~(\ref{eq:ML_det_exact}).
Here, 1-bit quantization is superior to multi-bit quantization over a relatively large \ac{SNR} region up to \SI{2.9}{dB}.
For higher \ac{SNR}, 2-bit quantization always yields the smallest \ac{SER}.
Hence, it seems to be not useful to increase the quantization resolution beyond $\log_2(M')$ bit.

\begin{figure}[t]
    \centering
    \subfloat[$b=1$, $N=64$.]{\includegraphics[width=\plotwidth\textwidth]{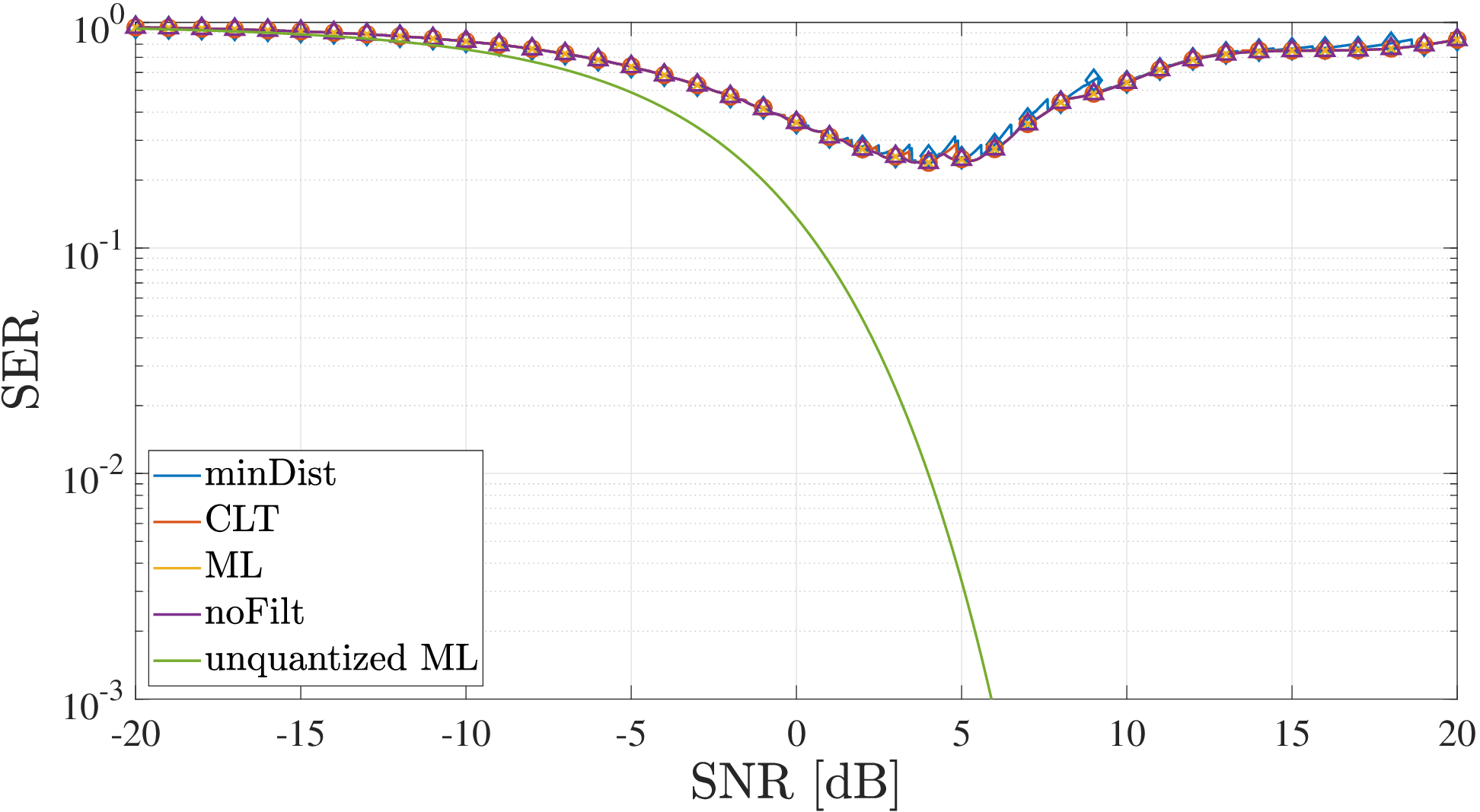}
        \label{fig:SER_1-bit_N64_8-ASK_detectors}}\\
    \subfloat[$b=2$, $N=32$, $\Delta=4$.]{\includegraphics[width=\plotwidth\textwidth]{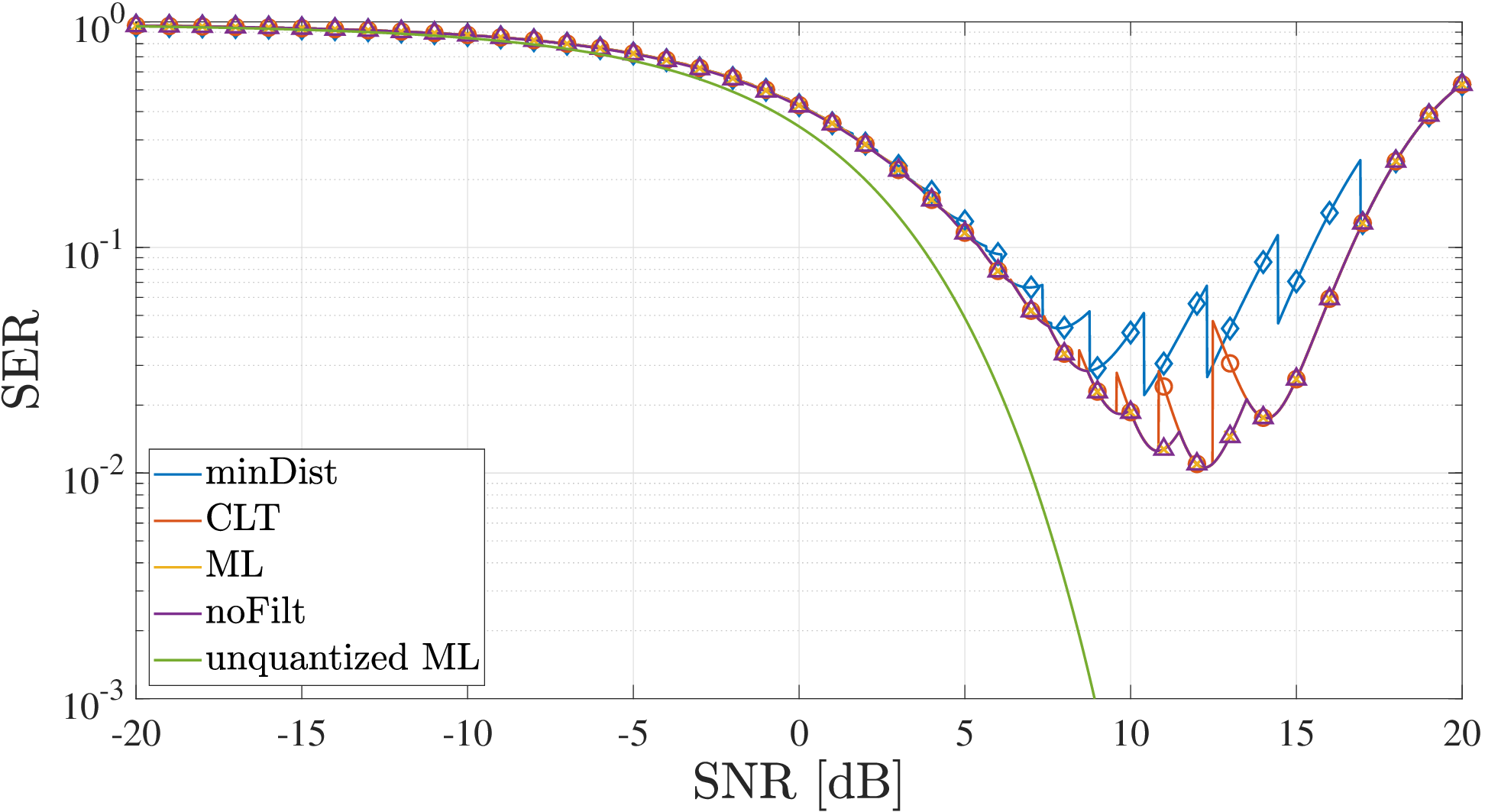}
        \label{fig:SER_2-bit_N32_8-ASK_detectors}}\\
    \subfloat[$b=3$, $N=16$, $\Delta=2$.]{\includegraphics[width=\plotwidth\textwidth]{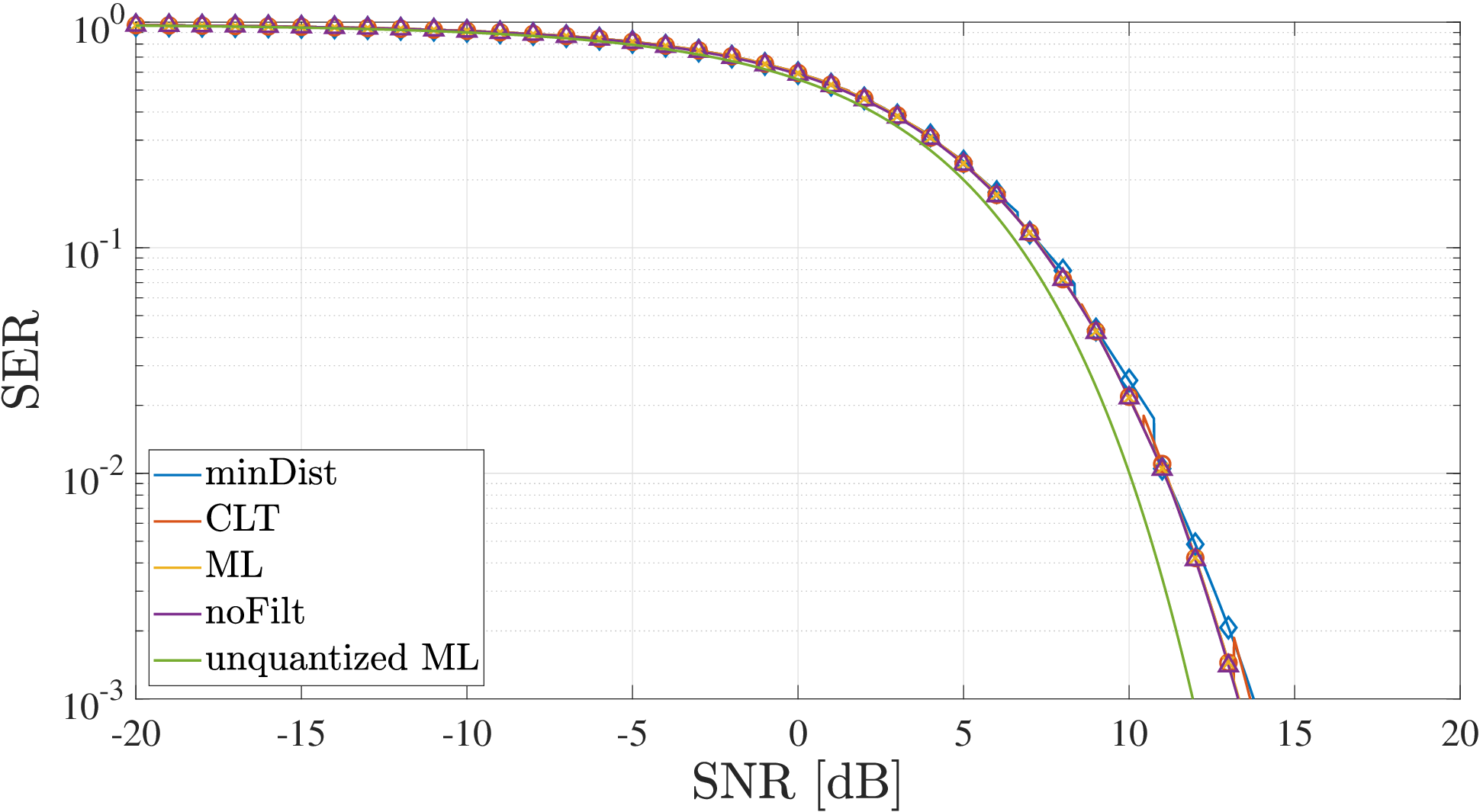}
        \label{fig:SER_3-bit_N16_8-ASK_detectors}}
    \caption{Analytical \acs{SER} (solid lines) and simulated \ac{SER} (markers) for the \acs{ML} detector~(\ref{eq:ML_det_exact}), CLT-based detector~(\ref{eq:ML_det_CLT}), minimum distance detector~(\ref{eq:subopt_det}), and fully optimum detector without filtering~(\ref{eq:z_ML_det}) under 64-\acs{QAM} transmission with $\mathcal{X}'=\{\pm1,\pm3,\pm5,\pm7\}$. Furthermore, the \ac{ML} detection performance for the unquantized case with the same oversampling factor is shown.}
    \label{fig:SER_8-ASK_detectors}
\end{figure}

\begin{figure}[t]
    \centerline{\includegraphics[width=\plotwidth\textwidth]{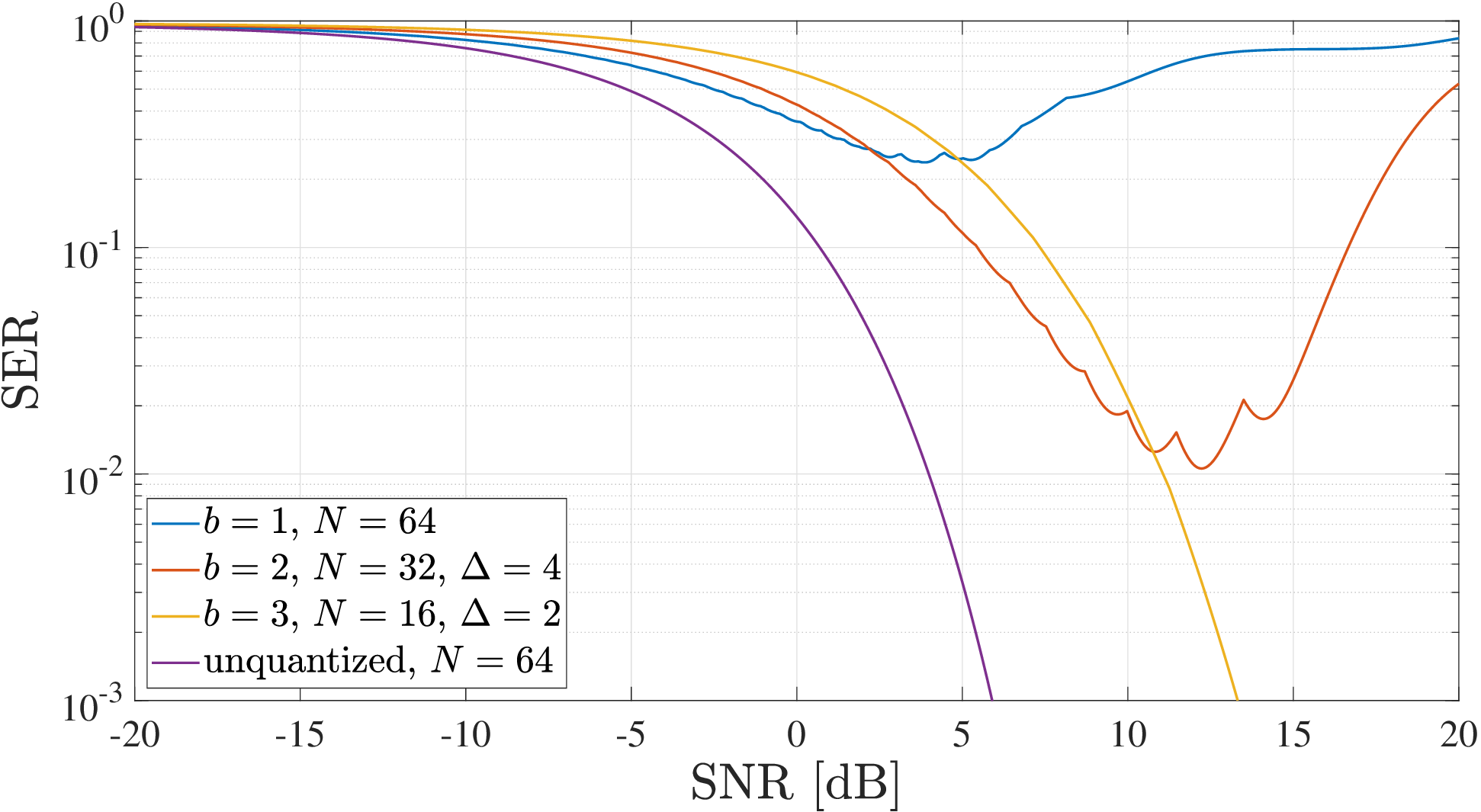}}
    \caption{Analytical {\acs{SER}} for the {\acs{ML}} detector~({\ref{eq:ML_det_exact}}) and different {\ac{ADC}} parametrizations under 64-{\ac{QAM}} transmission with $\mathcal{X}'=\{\pm1,\pm3,\pm5,\pm7\}$.}
    \label{fig:SER_1-3-bit_8-ASK_ML}
\end{figure}

Next, we analyze the \ac{SER} performance for a 64-\ac{QAM} constellation with $\mathcal{X}'=\{\pm1,\pm3,\pm5,\pm7\}$ in Fig.~\ref{fig:SER_8-ASK_detectors}.
The minimum \ac{SER} for 1-bit quantization is quite high and lies at approximately 0.24 due to the large constellation and low resolution.
Here, the performance could be enhanced by increasing the number of receive antennas $N$ or by employing optimized transmit symbol constellations as discussed in Section~{\ref{sec:const_opt}}.
The receiver with 2-bit quantization yields a minimum \ac{SER} of $\text{SER}=10^{-2}$.
Besides, the \ac{SER} increases for $b=1$ and $b=2$ at high \ac{SNR} because there are more constellation symbols than quantization levels.
For 3-bit quantization, the \ac{SER} decays to zero for high \ac{SNR} as expected.
Furthermore, we observe a performance gap between the optimal detector based on $\iq{d}$ and the two suboptimal detectors as in the 16-\ac{QAM} transmission case.
Moreover, it can be observed again that the analytically obtained \ac{SER} results match exactly with the \ac{SER} results obtained by Monte Carlo simulations.
Finally, when comparing the results for the different \ac{ADC} parametrizations in Fig.~\ref{fig:SER_1-3-bit_8-ASK_ML}, we deduce that 1-bit quantization performs best up to an \ac{SNR} of approximately \SI{2.2}{dB}.
Then, there is an \ac{SNR} window up to approximately \SI{10.8}{dB} for which 2-bit quantization yields the minimum \ac{SER}.
However, for $\text{SNR}>\SI{10.8}{dB}$, 3-bit quantization performs best.
For both 16-\ac{QAM} and 64-\ac{QAM} transmission, at low \acp{SNR} below a certain threshold, 1-bit quantization yields the smallest \ac{SER} when constraining the \ac{ADC} power consumption to be constant.

\subsection{Realistic Indoor THz Channel}\label{subsec:THz}
In this subsection, we consider a realistic frequency-selective indoor \ac{THz} channel with multipath propagation including reflected and scattered rays.
We adopt a corresponding ray tracing based channel model~\cite{Moldovan2014} which considers spherical wave propagation in the near field.
We show that our proposed \ac{ML} detector~\eqref{eq:ML_det_exact} performs similarly well for this realistic scenario as for the simplified scenario of a frequency-flat single-path \ac{LoS} channel, analyzed in Subsection~\ref{subsec:AWGN}.
We consider data transmission at a carrier frequency of \SI{300}{GHz} with a bandwidth of \SI{20}{GHz} in a large rectangular room with a length, width, and height of \SI{10}{m}, \SI{20}{m}, and \SI{2.5}{m}, respectively.
The walls of the room are made of plaster "sample s2" from~\cite{Piesiewicz2007} which has a high roughness.
We assume a transmit power of \SI{13}{dBm}~\cite{John2020} and employ a \ac{RRC} filter with a roll-off factor of 0.25 at both transmitter and receiver side.
This results in a data rate of \SI{64}{Gbps} and \SI{96}{Gbps} for 16- and 64-\ac{QAM} transmission, respectively.
Furthermore, we consider perfectly aligned horn antennas at both the transmitter and the receiver.
Here, we use the Gaussian beam model for the radiation pattern of the horn antenna with a gain of \SI{18.9}{dBi}~\cite{Priebe2012}.
Besides, we utilize a \ac{UPA} with $\lambda/2$ antenna spacing at the receiver with $8\times8$, $8\times4$, and $4\times4$ antennas in the horizontal and vertical direction for $N=64$, $N=32$, and $N=16$, respectively.
Finally, the noise power density is set to \SI{-174}{dBm/Hz}, and a noise figure of \SI{15}{dB} is selected.
The simulation parameters are summarized in Table~\ref{tab:sim_parameters}.
\begin{table}[tb]
\caption{\Ac{THz} simulation parameters.}
\begin{center}
\begin{tabular}{|c|c|}
\hline
\textbf{Parameter}&\textbf{Value} \\
\hline
Carrier frequency&\SI{300}{GHz} \\
\hline
Bandwidth&\SI{20}{GHz} \\
\hline
Distance&\SIrange[]{1}{17}{m} \\
\hline
Wall material&Plaster \\
\hline
Transmit power&\SI{13}{dBm} \\
\hline
Transmit pulse and receive filter&\Ac{RRC} filter with roll-off factor 0.25 \\
\hline
Transmit and receive antennas&Horn antenna with \SI{18.9}{dBi} gain \\
\hline
Noise power density&\SI{-174}{dBm/Hz} \\
\hline
Noise figure&\SI{15}{dB} \\
\hline
\end{tabular}
\label{tab:sim_parameters}
\end{center}
\end{table}

\begin{figure}[t]
    \centerline{\includegraphics[width=\plotwidth\textwidth]{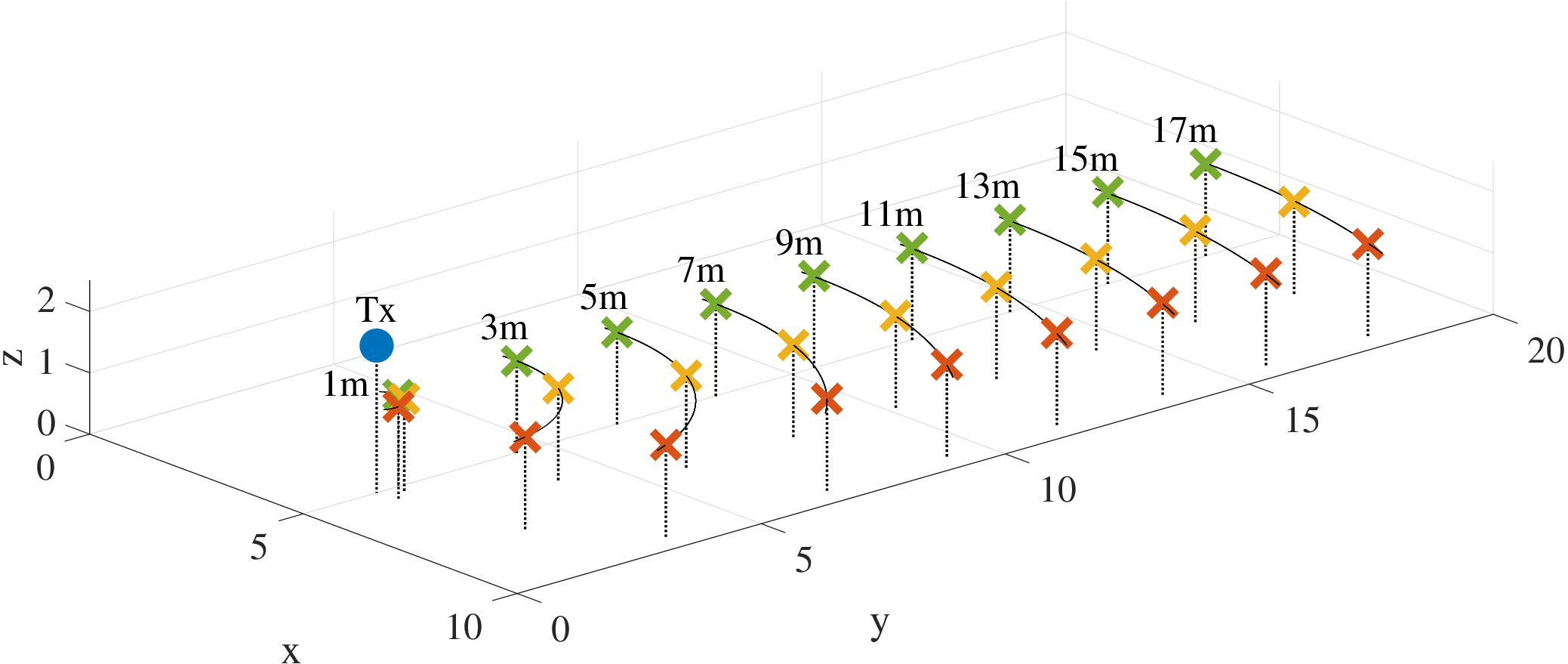}}
    \caption{Indoor environment with transmitter and receiver positions.}
    \label{fig:TX_RX_positions}
\end{figure}

The transmitter is placed at one side of the room at the coordinates \SI{5}{m}, \SI{1.5}{m}, \SI{2.4}{m} as illustrated in Fig.~\ref{fig:TX_RX_positions}.
The receiver is located at a height of \SI{1.5}{m} at nine different distances from the transmitter, ranging from \SI{1}{m} to \SI{17}{m}.
For each distance, we consider three different positions of the receiver: one in the middle of the room, one close to the wall, and one in between as can be also observed from Fig.~\ref{fig:TX_RX_positions}.
For each receiver position, we transmit 100 independent blocks, each consisting of $10^4$ consecutive symbols.
\ac{ISI} occurs only between the symbols within one block due to protection intervals between the blocks.
Then, the \ac{SER} results corresponding to the three different positions with a given distance are averaged.
Hence, we consider in total $3\cdot10^6$ symbol tranmissions for each distance.
Finally, in order to guarantee a good detection performance even for small distances, i.e., for high \ac{SNR}, we add artificial noise in front of the \acp{ADC} for 16-\ac{QAM} and 64-\ac{QAM} transmission in case of 1-bit quantization and for 64-\ac{QAM} transmission in case of 2-bit quantization, respectively.
The optimum power of the artificial noise is determined by inspecting the analytical results from Subsection~\ref{subsec:AWGN} and identifying the \ac{SNR} value which yields the smallest \ac{SER}.
According to Figs.~\ref{fig:SER_1-bit_N64_4-ASK_detectors} and~\ref{fig:SER_1-bit_N64_8-ASK_detectors}, this optimum \ac{SNR} value is given by approximately \SI{5.6}{dB} and \SI{3.8}{dB} for 16-\ac{QAM} and 64-\ac{QAM} transmission, respectively, in case of 1-bit quantization.
For 2-bit quantization and 64-\ac{QAM}, the optimum \ac{SNR} value is \SI{12.2}{dB} according to Fig.~\ref{fig:SER_2-bit_N32_8-ASK_detectors}.
Note that these optimum \ac{SNR} values can be only reached when the actual \ac{SNR} is larger than the optimum value which is the case for relatively small distances.
When the actual \ac{SNR} is lower than the optimum value, i.e., for large distances, we cannot enhance the performance by adding artificial noise.

\begin{figure}[t]
    \centering
    \subfloat[16-{\ac{QAM}} transmission with $\mathcal{X}'=\{\pm1,\pm3\}$.]{\includegraphics[width=\plotwidth\textwidth]{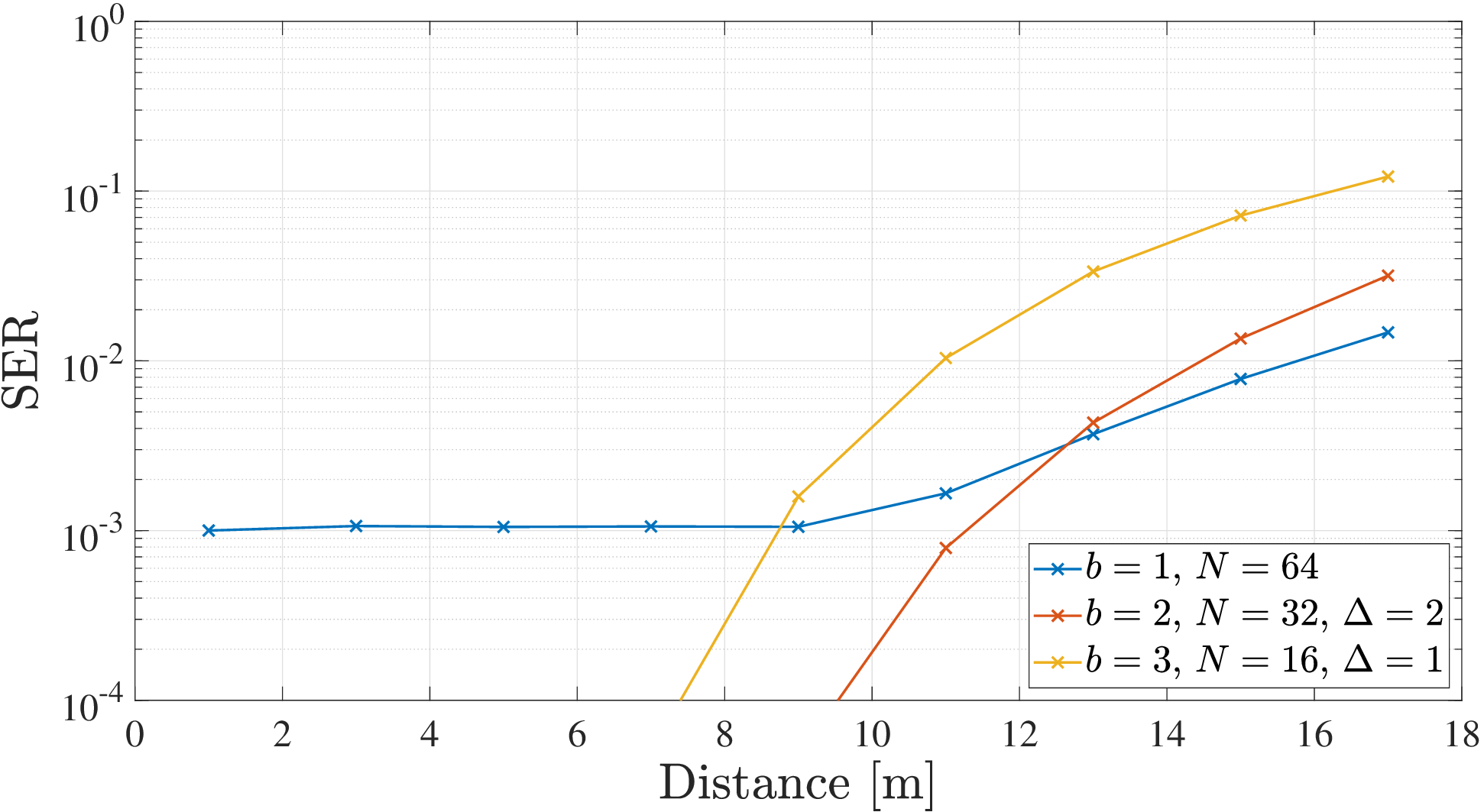}
    \label{fig:SER_THz_1-3-bit_4-ASK_ML}}\\
    \subfloat[64-{\ac{QAM}} transmission with $\mathcal{X}'=\{\pm1,\pm3,\pm5,\pm7\}$.]{\includegraphics[width=\plotwidth\textwidth]{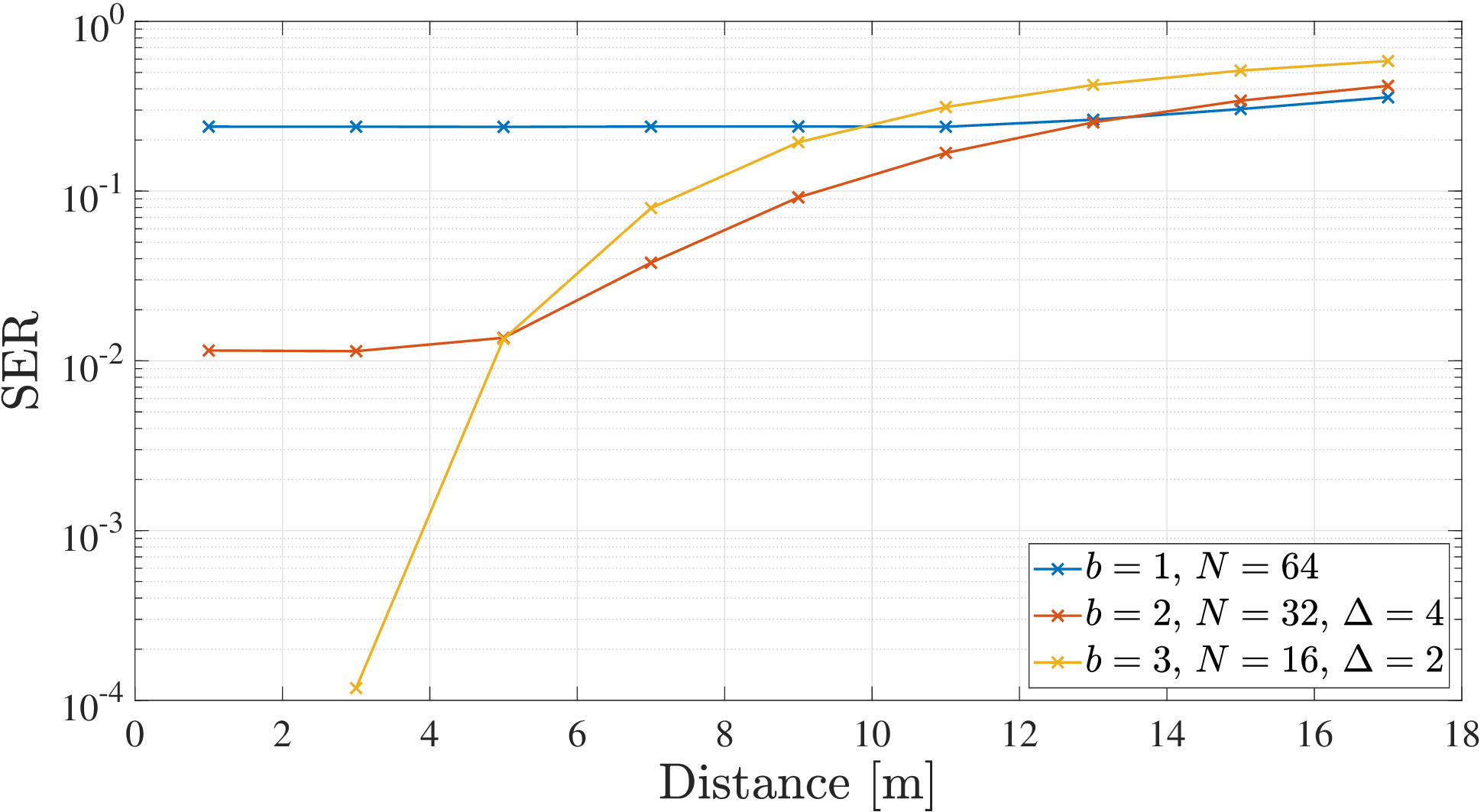}
    \label{fig:SER_THz_1-3-bit_8-ASK_ML}}\\
    \caption{Simulated {\acs{SER}} versus distance in a realistic indoor {\ac{THz}} channel for the {\acs{ML}} detector~({\ref{eq:ML_det_exact}}) and different {\ac{ADC}} parametrizations.}
    \label{fig:SER_THz_1-3-bit_ML}
\end{figure}

In Fig.~\ref{fig:SER_THz_1-3-bit_4-ASK_ML}, \ac{SER} versus the communication distance is shown for 16-\ac{QAM}.
Here, we can observe that 2-bit quantization performs best at small distances.
This is in agreement with the results in Fig.~\ref{fig:SER_1-3-bit_4-ASK_ML} for high \ac{SNR}.
For distances larger than \SI{13}{m}, 1-bit quantization performs best.
Furthermore, we observe that adding artificial noise before 1-bit quantization enhances the performance for small distances and keeps the error rate constant up to a distance of approximately \SI{9}{m}.
In this case, the \ac{SER} is as small for sufficiently small distances as for the simplified frequency-flat single-path \ac{LoS} channel in Fig.~\ref{fig:SER_1-bit_N64_4-ASK_detectors}.
Hence, we conclude that the simplifying assumptions are justified for the considered indoor THz channel, i.e., \ac{ISI} can be neglected, and all channel gains are approximately equal.
Hence, our proposed detector is still (close-to-)optimal in this case.
\ac{SER} versus distance for 64-\ac{QAM} transmission is shown in Fig.~\ref{fig:SER_THz_1-3-bit_8-ASK_ML}.
Here, similar observations can be made as in case of 16-\ac{QAM} before.
1- and 2-bit quantization benefit from adding artificial noise, whereas 3-bit quantization only performs best at small distances. A comparison with the results of Fig.~\ref{fig:SER_8-ASK_detectors} again confirms the validity of the simplifying assumptions adopted for detector design.

It should be noted that a similar performance as for small distances can be also attained for larger distances by increasing the \ac{SNR} via, e.g., increasing the transmit power, using more directional antennas, or decreasing the noise level at the receiver.

\begin{figure}[t]
    \centering
    \subfloat[16-{\ac{QAM}} transmission with $\mathcal{X}'=\{\pm1,\pm3\}$.]{\includegraphics[width=\plotwidth\textwidth]{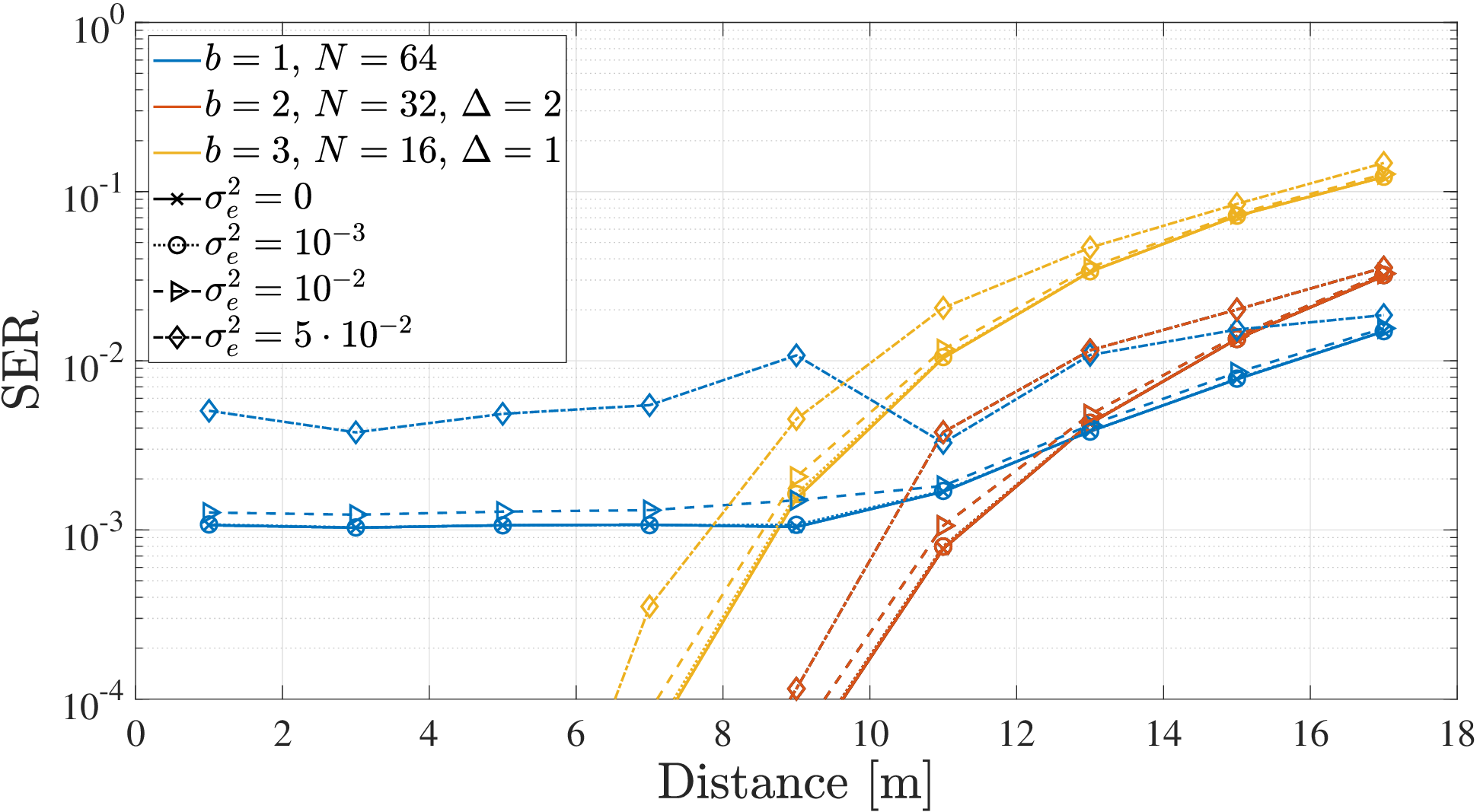}
    \label{fig:SER_THz_1-3-bit_4-ASK_ML_chErr}}\\
    \subfloat[64-{\ac{QAM}} transmission with $\mathcal{X}'=\{\pm1,\pm3,\pm5,\pm7\}$.]{\includegraphics[width=\plotwidth\textwidth]{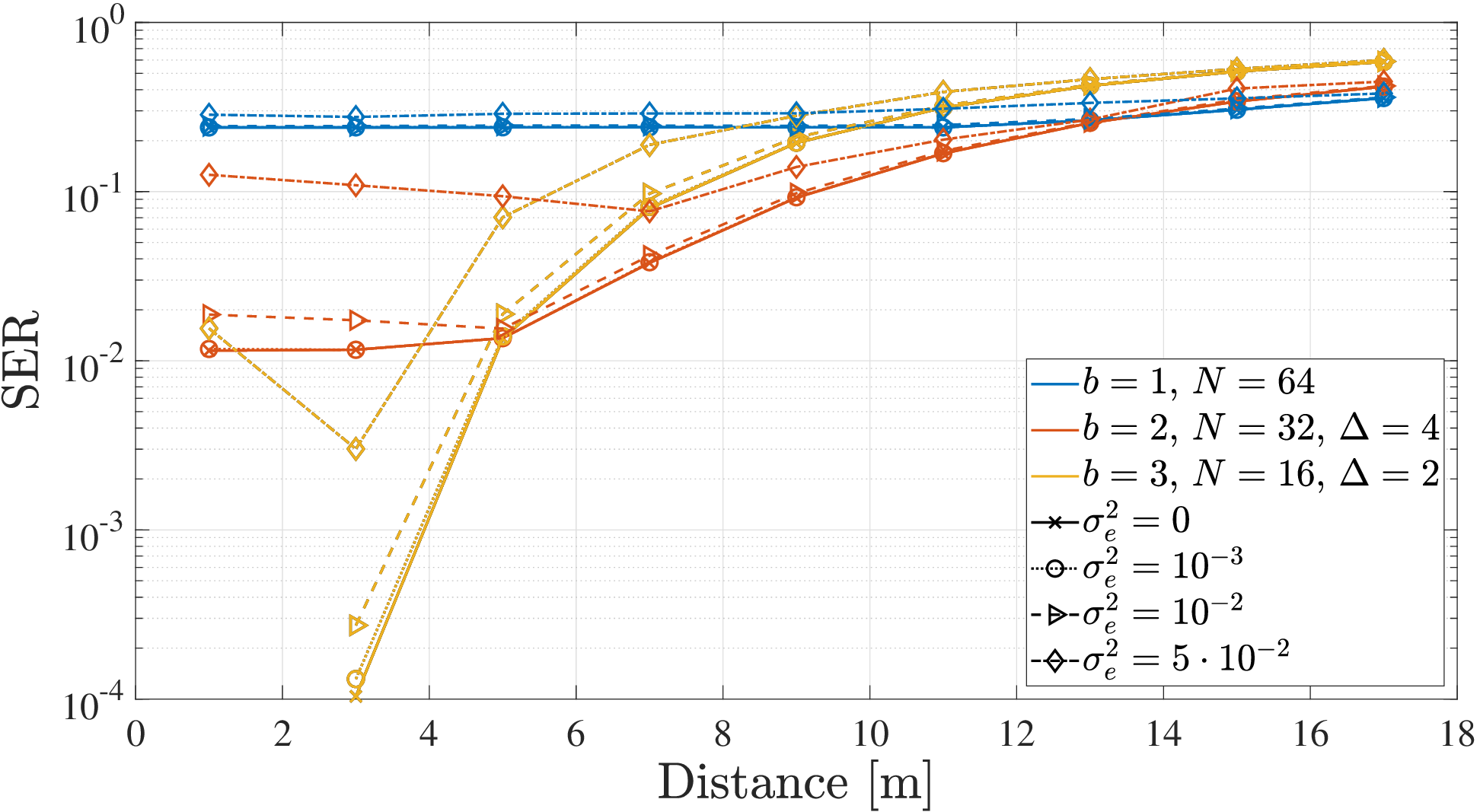}
    \label{fig:SER_THz_1-3-bit_8-ASK_ML_chErr}}\\
    \caption{Simulated {\acs{SER}} versus distance in a realistic indoor {\ac{THz}} channel with channel estimation errors for the {\acs{ML}} detector~({\ref{eq:ML_det_exact}}) and different {\ac{ADC}} parametrizations.}
    \label{fig:SER_THz_1-3-bit_ML_chErr}
\end{figure}

For the previous results, we have assumed perfect knowledge of the {\ac{LoS}} channel coefficients for detection.
In the following, we show numerical results which include the effect of channel estimation errors.
We consider a zero-mean additive white Gaussian error for the estimated channel coefficients with variance $\sigma_e^2$ with respect to the {\ac{LoS}} power, i.e., the error power is equal to the power of the {\ac{LoS}} channel coefficients multiplied by $\sigma_e^2$.
Such modeling of the statistics of the channel estimation errors is accurate for pilot-based channel estimation with optimized pilot sequences.
The corresponding results are illustrated in Fig.~{\ref{fig:SER_THz_1-3-bit_ML_chErr}}.
It can be observed that there is essentially no performance degradation for sufficiently small channel estimation errors, i.e., $\sigma_e^2=10^{-3}$.
For medium channel estimation errors, $\sigma_e^2=10^{-2}$, there is a small but negligible performance degradation.
Only for strong channel estimation errors with a power of $\sigma_e^2=5\cdot10^{-2}$ below {\ac{LoS}}, there is a noticeable loss in performance which is still acceptable for certain scenarios.
Overall, the results show that our scheme is able to cope with small-to-moderate channel estimation errors which can be guaranteed in a practical system by employing a sufficiently long pilot sequence for channel estimation.

\section{Optimization of Higher-Order Constellations}\label{sec:const_opt}
One main difference of the statistics of the detection variable $\iq{d}$ compared to those of the detection variable $\iq{y}=\iq{x}+\iq{n}$ of an unquantized Nyquist-rate sampling transmission is that the variance of $\iq{d}$ depends on the transmitted symbol, whereas the variance of $\iq{y}$ is identical for all input symbols.
Due to the equal variance in the unquantized case, it is optimal to place the input symbols equidistantly, i.e., $|\iq{x}_1-\iq{x}_2|=|\iq{x}_2-\iq{x}_3|=\ldots=|\iq{x}_{M'-1}-\iq{x}_{M'}|$.
However, for the quantized system, this is not valid anymore.
Especially for quantized systems with higher-order constellations, the variance can vary significantly for different input symbols as can be seen exemplarily for 1-bit quantization in Fig.~\ref{fig:d_distr_1bit_4ASK}.
Therefore, it is not obvious what is the optimum constellation corresponding to minimum \ac{SER}.
In the following, we aim at determining the optimum higher-order constellations for 1-bit and 2-bit quantization.

For 1-bit quantization, we focus on symmetric square 16-, 36-, and 64-\ac{QAM} constellations with one, two, and three degrees of freedom, respectively, for constellation optimization.
Here, the inner four symbols are fixed to $\pm1\pm\imag$, whereas the other symbols are variable but symmetrically arranged.
Therefore, the quadrature component constellations are given by $\mathcal{X}'=\{\pm1,\pm a_1\}$ for 16-\ac{QAM}, $\mathcal{X}'=\{\pm1,\pm a_1,\pm a_2\}$ for 36-\ac{QAM}, and $\mathcal{X}'=\{\pm1,\pm a_1,\pm a_2,\pm a_3\}$ for 64-\ac{QAM}.
For 2-bit quantization, we consider symmetric square 64-{\ac{QAM}} constellations with four degrees of freedom.
Here, there is one more degree of freedom compared to the respective constellation for 1-bit quantization since in multi-bit quantization we have an additional degree of freedom represented by the quantization step size $\Delta$.
For the following, we choose to fix the step size $\Delta$ and optimize all levels of our constellation, i.e., the quadrature component constellation is given by $\mathcal{X}'=\{\pm a_1,\pm a_2,\pm a_3,\pm a_4\}$ with four degrees of freedom.

We find the optimal constellations empirically in each case by comparing \ac{SER} for different choices of the available free parameters.
Although our investigations are limited to 1-bit and 2-bit quantization, a similar procedure can be carried out for any quantization resolution.
We note that an analytical optimization appears intractable because of the highly nonlinear nature of the problem.

\subsection{1-Bit Quantization}\label{subsec:1bit_const_opt}
\begin{figure}[t]
    \centering
    \subfloat[Analytical {\acs{SER}} versus {\ac{SNR}}.]{\includegraphics[width=\plotwidth\textwidth]{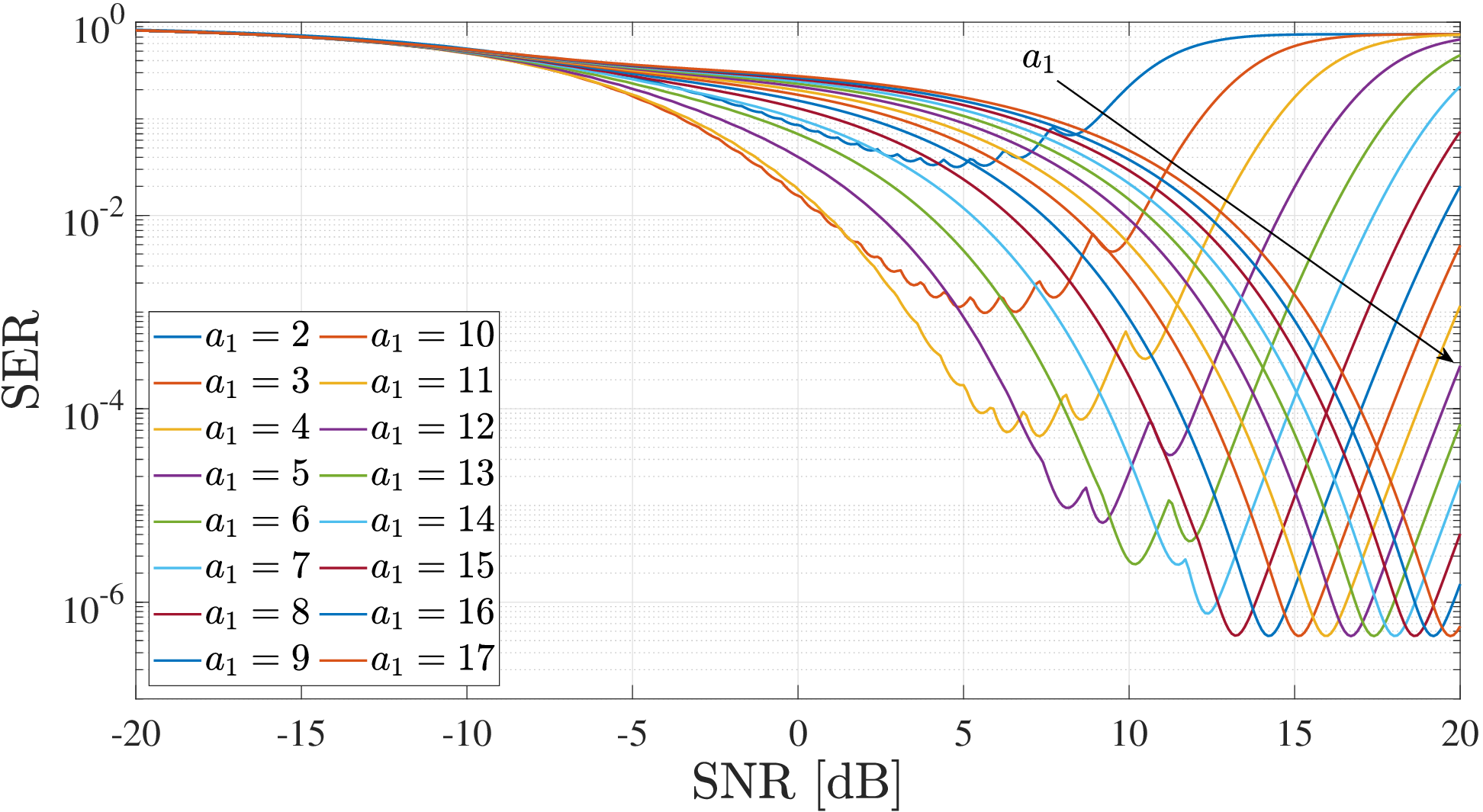}
    \label{fig:SER_1-bit_N64_4-ASK_ML_opt_const}}\\
    \subfloat[Minimum achievable {\acs{SER}} versus constellation parameter $a_1$.]{\includegraphics[width=\plotwidth\textwidth]{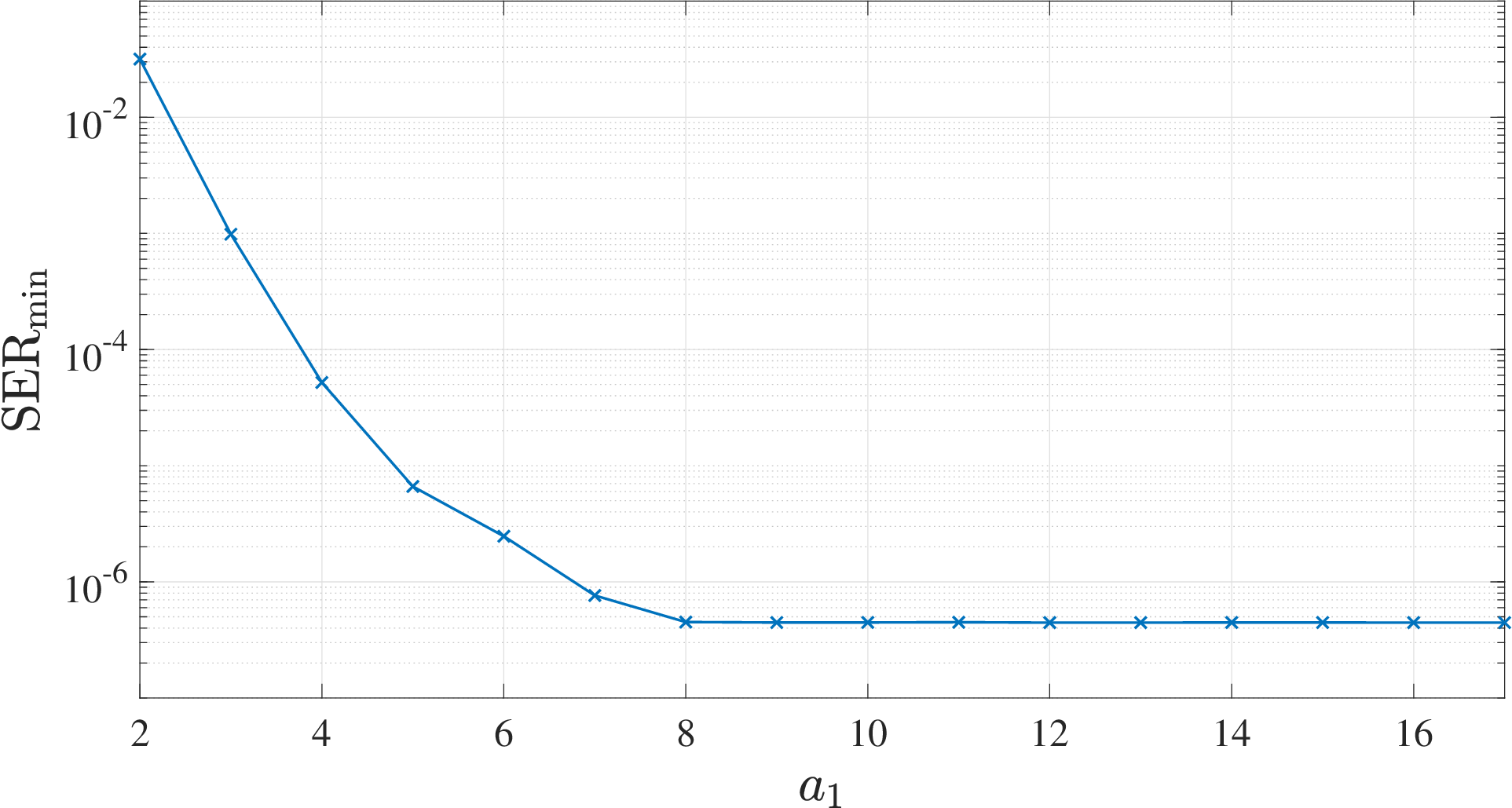}
    \label{fig:SER_min_1-bit_N64_4-ASK_ML_opt_const}}\\
    \caption{{\acs{SER}} for the {\acs{ML}} detector~({\ref{eq:ML_det_exact}}) and different 16-{\acs{QAM}} constellations with $\mathcal{X}'=\{\pm1,\pm a_1\}$, $b=1$, and $N=64$.}
    \label{fig:1-bit_N64_4-ASK_ML_opt_const}
\end{figure}

In Fig.~\ref{fig:SER_1-bit_N64_4-ASK_ML_opt_const}, analytically obtained \ac{SER} curves are shown for the case of 1-bit quantization with the \ac{ML} detector~(\ref{eq:ML_det_exact}) and various 16-\ac{QAM} constellations with different values of $a_1$.
It can be observed that the minimum \ac{SER} can be decreased significantly by increasing $a_1$.
For example, the minimum \ac{SER} for $a_1=3$ is $9.8\cdot10^{-4}$, while the constellation with $a_1=8$ yields a minimum \ac{SER} of $4.5\cdot10^{-7}$.
This emphasizes the importance of choosing an optimized constellation.
However, the minimum \ac{SER} for $a_1=8$ is achieved at a higher \ac{SNR} than that for $a_1=3$.
Another interesting observation can be made for $a_1>8$.
Here, the location of the minimum \ac{SER} moves to higher \ac{SNR} for increasing $a_1$ while the minimum \ac{SER} value itself does not decrease anymore.
This suggests that the minimum \ac{SER}, illustrated in Fig.~\ref{fig:SER_min_1-bit_N64_4-ASK_ML_opt_const}, converges to a fixed value for increasing $a_1$, and $a_1\approx8$ is the smallest outer quadrature component constellation point for which this value can be reached closely.
However, the outer constellation point with this property depends on the oversampling factor.
Hence, the constellation optimization has to be carried out individually for each oversampling factor of interest.
Furthermore, we note that the results indicate that for increasing \ac{SNR} the optimal outer points will become larger or equivalently the inner points shrink when including a power normalization into the constellation.
For $\text{SNR}\to\infty$, this will result in a 9-\ac{QAM} constellation which was also proven to be optimal in~\cite{Krone2012}.
Some further results on optimized 16-QAM constellations can be found in~\cite{Forsch2022}.

\begin{figure}[t]
    \centering
    \subfloat[Analytical {\acs{SER}} versus {\ac{SNR}}.]{\includegraphics[width=\plotwidth\textwidth]{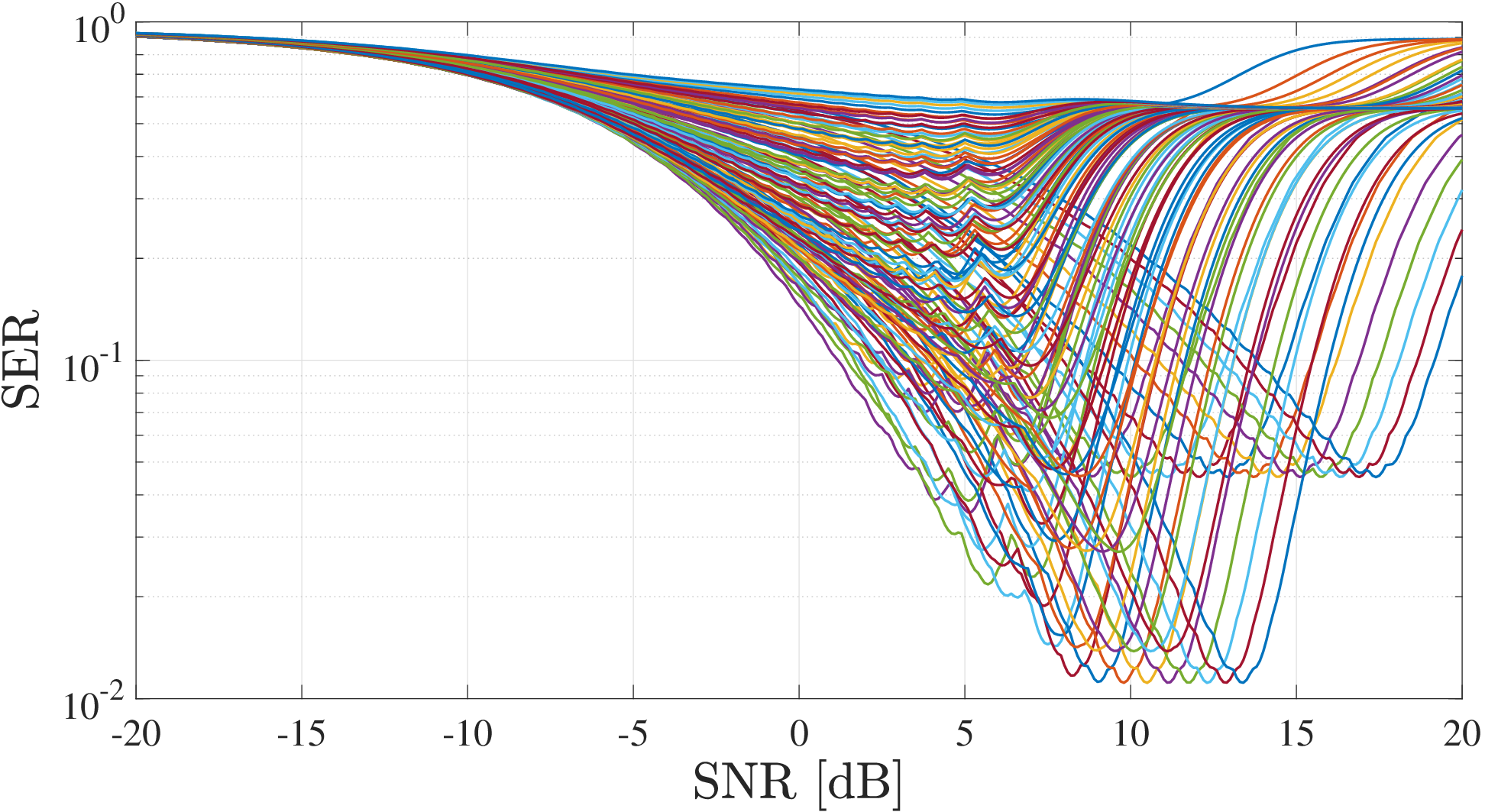}
    \label{fig:SER_1-bit_N64_6-ASK_ML_opt_const}}\\
    \subfloat[Minimum achievable {\acs{SER}} versus constellation parameters $a_1$ and $a_2$.]{\includegraphics[width=\plotwidth\textwidth]{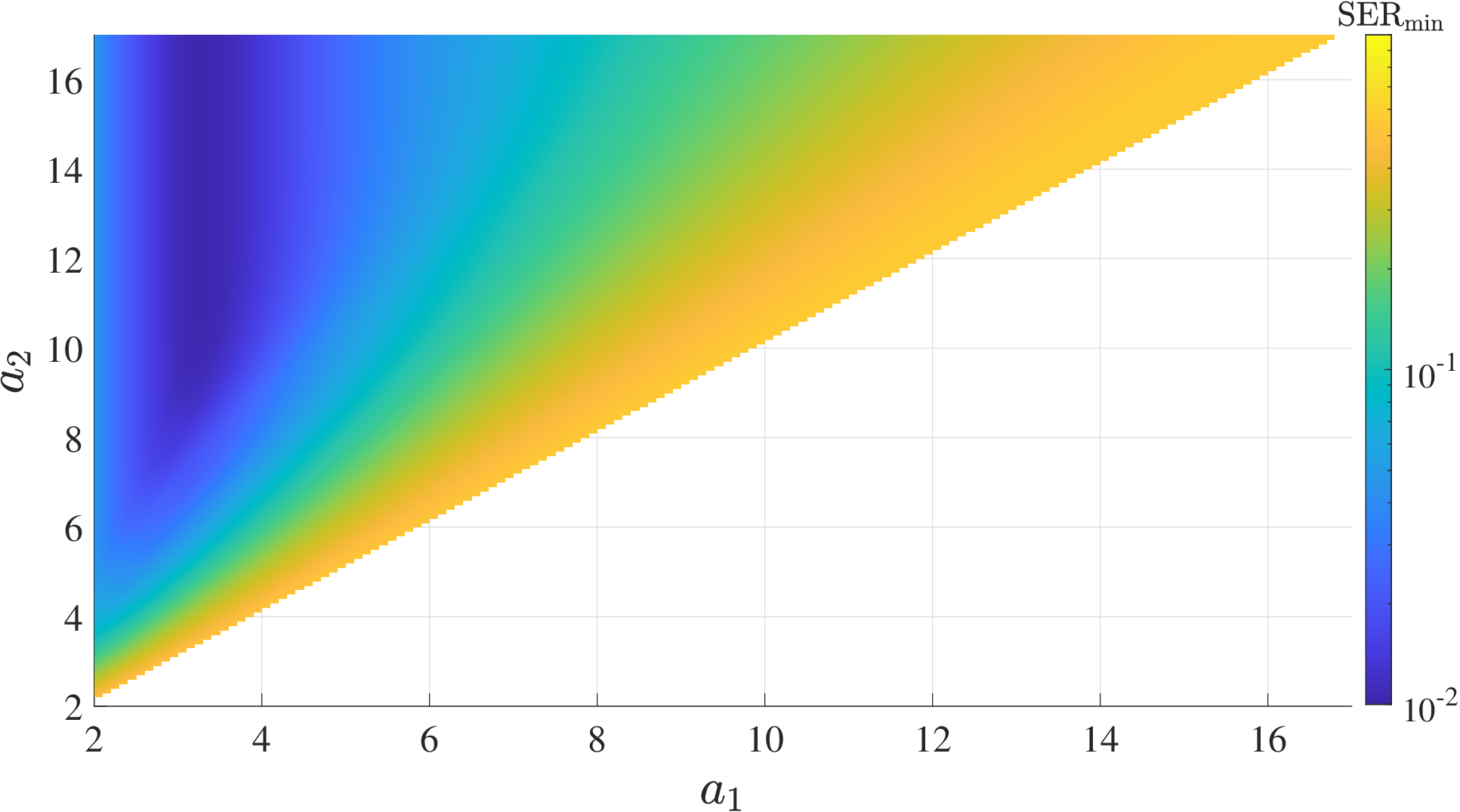}
    \label{fig:SER_min_1-bit_N64_6-ASK_ML_opt_const}}\\
    \caption{{\acs{SER}} for the {\acs{ML}} detector~({\ref{eq:ML_det_exact}}) and different 36-{\acs{QAM}} constellations with $\mathcal{X}'=\{\pm1,\pm a_1,\pm a_2\}$, $b=1$, and $N=64$.}
    \label{fig:1-bit_N64_6-ASK_ML_opt_const}
\end{figure}

Next, we investigate the \ac{SER} performance for different 36-\ac{QAM} constellations with $b=1$, illustrated in Fig.~\ref{fig:SER_1-bit_N64_6-ASK_ML_opt_const}.
Here, a large number of constellations is considered with all possible integer values for $a_1$ and $a_2$ with $2\leq a_1<a_2\leq17$.
As for the 16-\ac{QAM} constellations, we observe different minimum \ac{SER} values for different constellations, while multiple constellations yield the same minimum \ac{SER} at different \ac{SNR}.
The minimum \ac{SER} versus the constellation parameters $a_1$ and $a_2$ is shown in Fig.~\ref{fig:SER_min_1-bit_N64_6-ASK_ML_opt_const}.
Here, also non-integer values of $a_1$ and $a_2$ have been included.
From the analytically obtained \ac{SER} values, it can be observed that, similarly to the 16-\ac{QAM} case, the minimum \ac{SER} converges for $a_1=3.3$ and $a_2>11$.
Hence, we can conclude that the position of the optimum middle quadrature component constellation point is almost identical to the classical case with $a_1=3$, but the position of the outer constellation point is moved to much higher values compared to the standard constellation.
Also, we can observe from Fig.~\ref{fig:SER_min_1-bit_N64_6-ASK_ML_opt_const} that selecting a large value for $a_1$ or choosing $a_2$ close to $a_1$ always results in a poor performance.
For the classical constellation with $a_1=3$ and $a_2=5$, a minimum \ac{SER} of $8.4\cdot10^{-2}$ is achieved, while for the optimized constellation with $a_1=3.3$ and $a_2>11$ a \ac{SER} of approximately $1.0\cdot10^{-2}$ cannot be undercut.

\begin{figure}[t]
    \centering
    \subfloat[Analytical {\acs{SER}} versus {\ac{SNR}}.]{\includegraphics[width=\plotwidth\textwidth]{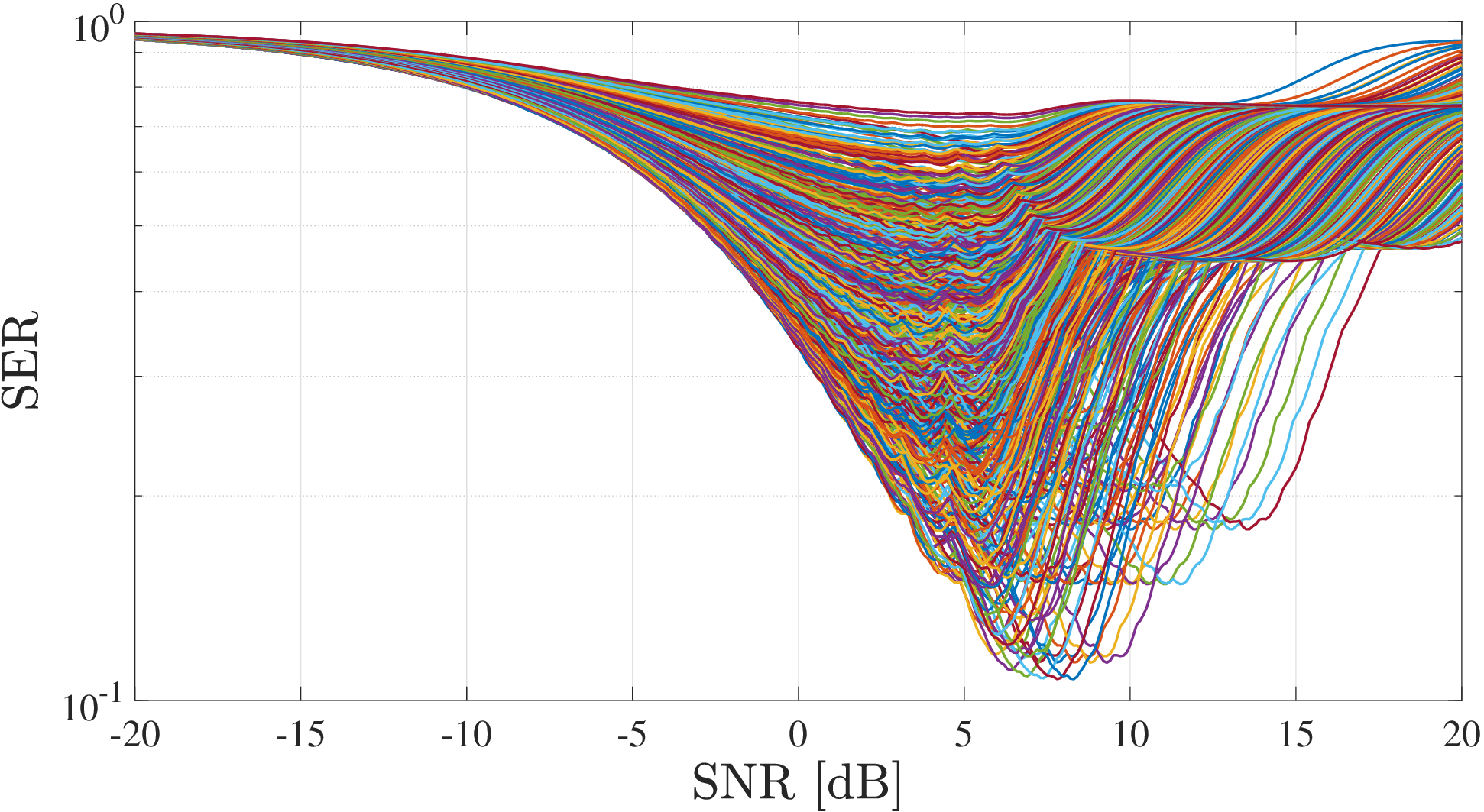}
    \label{fig:SER_1-bit_N64_8-ASK_ML_opt_const}}\\
    \subfloat[Minimum achievable {\acs{SER}} versus constellation parameters $a_1$, $a_2$, and $a_3$.]{\includegraphics[trim={19.3mm 2mm 0mm 0mm},clip,width=\plotwidth\textwidth]{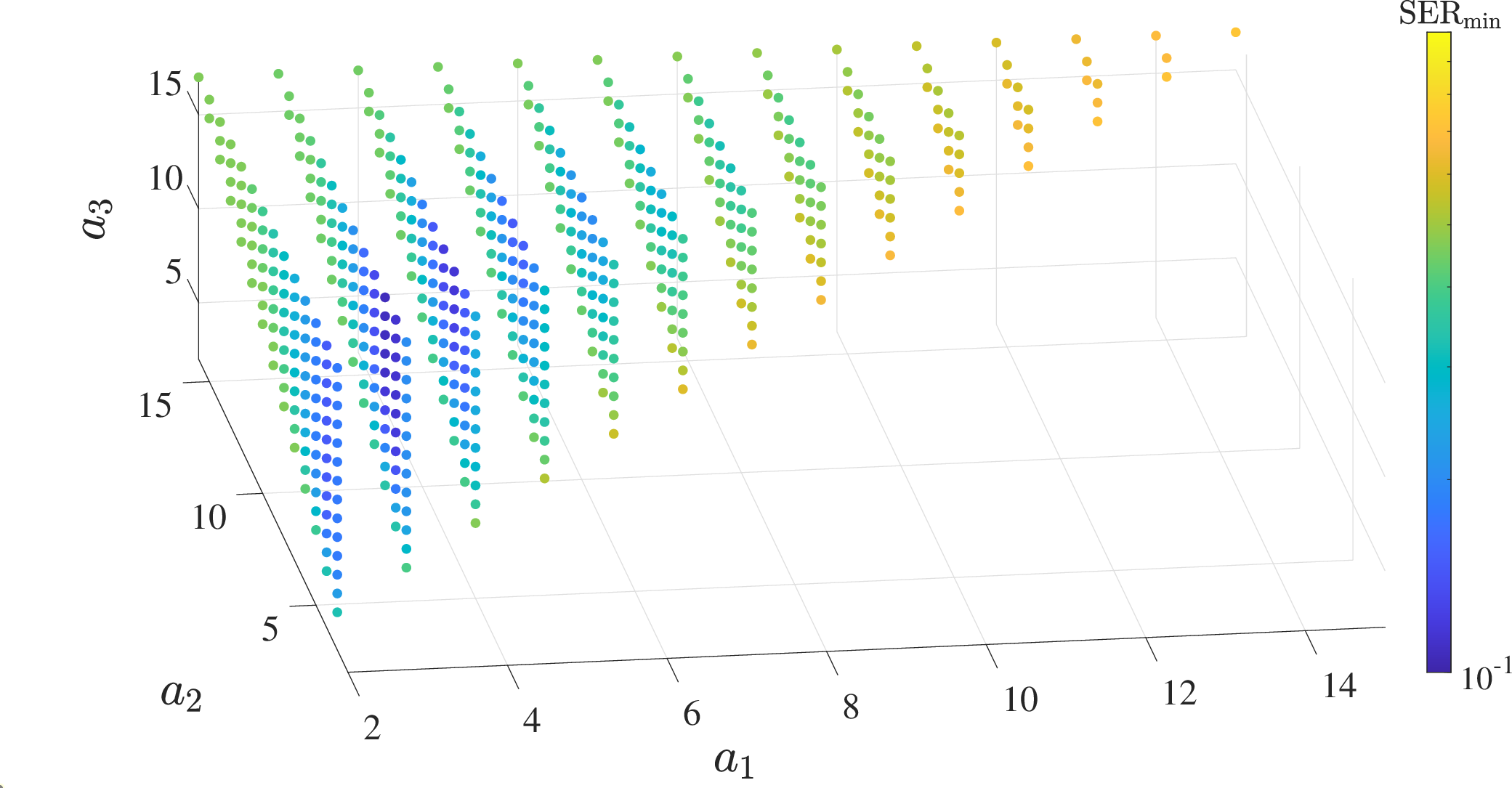}
    \label{fig:SER_min_1-bit_N64_8-ASK_ML_opt_const}}\\
    \caption{{\acs{SER}} for the {\acs{ML}} detector~({\ref{eq:ML_det_exact}}) and different 64-{\acs{QAM}} constellations with $\mathcal{X}'=\{\pm1,\pm a_1,\pm a_2,\pm a_3\}$, $b=1$, and $N=64$.}
    \label{fig:1-bit_N64_8-ASK_ML_opt_const}
\end{figure}

Now, we study for $b=1$ 64-\ac{QAM} constellations with three degrees of freedom, $a_1$, $a_2$, and $a_3$.
The corresponding \ac{SER} versus \ac{SNR} is depicted in Fig.~\ref{fig:SER_1-bit_N64_8-ASK_ML_opt_const} for different constellations with integer constellation points with $2\leq a_1<a_2<a_3\leq17$.
A similar behavior as for 16- and 36-\ac{QAM} constellations can be recognized.
The same holds for the minimum \ac{SER} which is illustrated in Fig.~\ref{fig:SER_min_1-bit_N64_8-ASK_ML_opt_const} for integer values of $a_1$, $a_2$, and $a_3$ for simplicity.
However, we also conducted a full search over a finer grid, showing that a minimum \ac{SER} of approximately $1.0\cdot10^{-1}$ is achieved for $a_1=3.1$, $a_2=5.7$, and $a_3>14$, whereas the classical constellation yields a minimum \ac{SER} of $2.4\cdot10^{-1}$.

\begin{figure}[t]
    \centering
    \subfloat[Classical 16-\acs{QAM}.]{\includegraphics[width=\plotwidthh\textwidth]{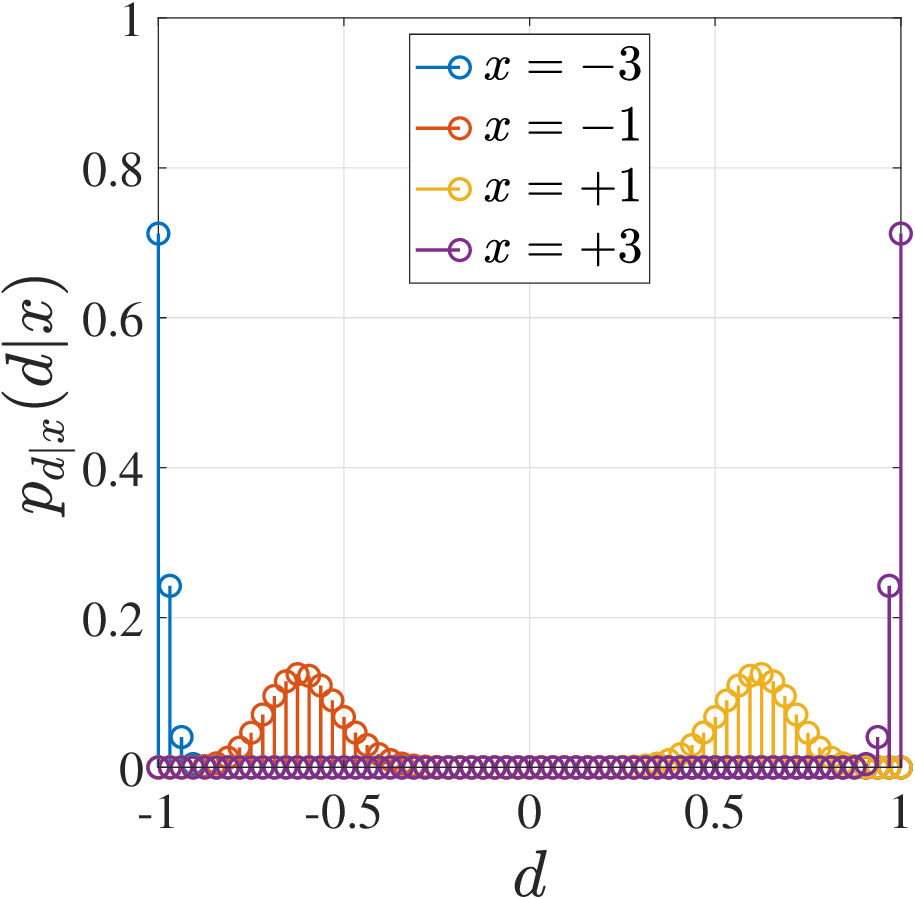}
        \label{fig:d_distr_1bit_4ASK_clas_const_opt_SNR}}
    \hfil
    \subfloat[Optimized 16-\acs{QAM}.]{\includegraphics[width=\plotwidthh\textwidth]{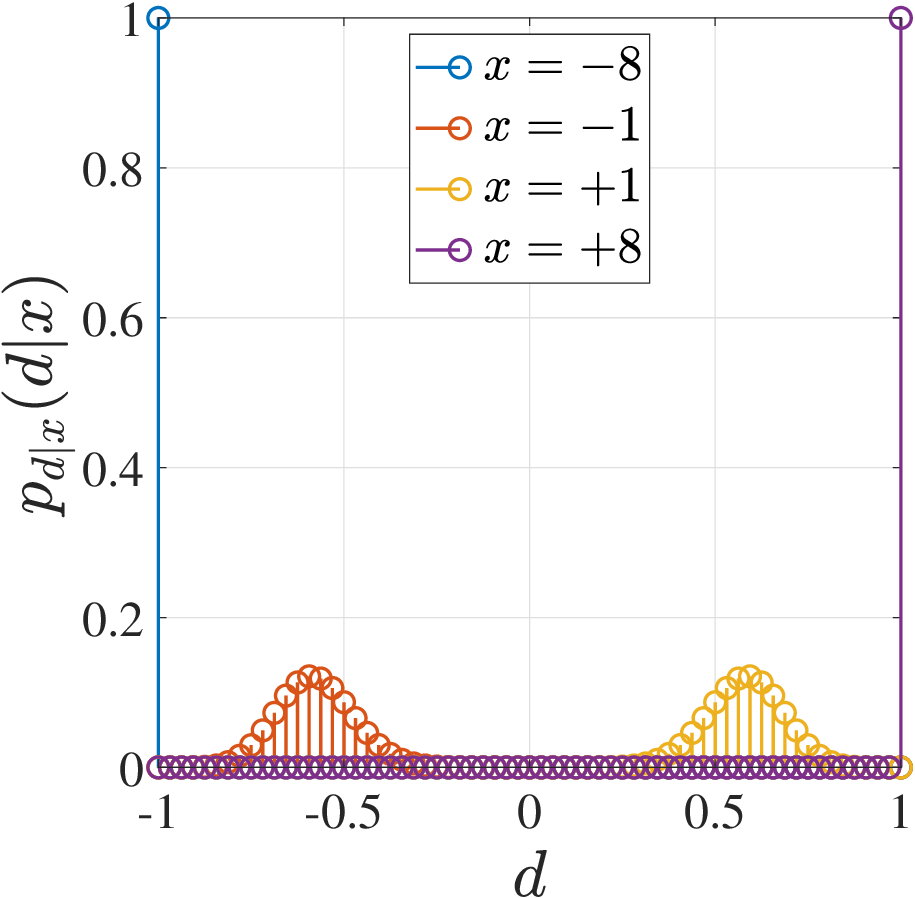}
        \label{fig:d_distr_1bit_4ASK_opt_const_opt_SNR}}\\
    \subfloat[Classical 64-\acs{QAM}.]{\includegraphics[width=\plotwidthh\textwidth]{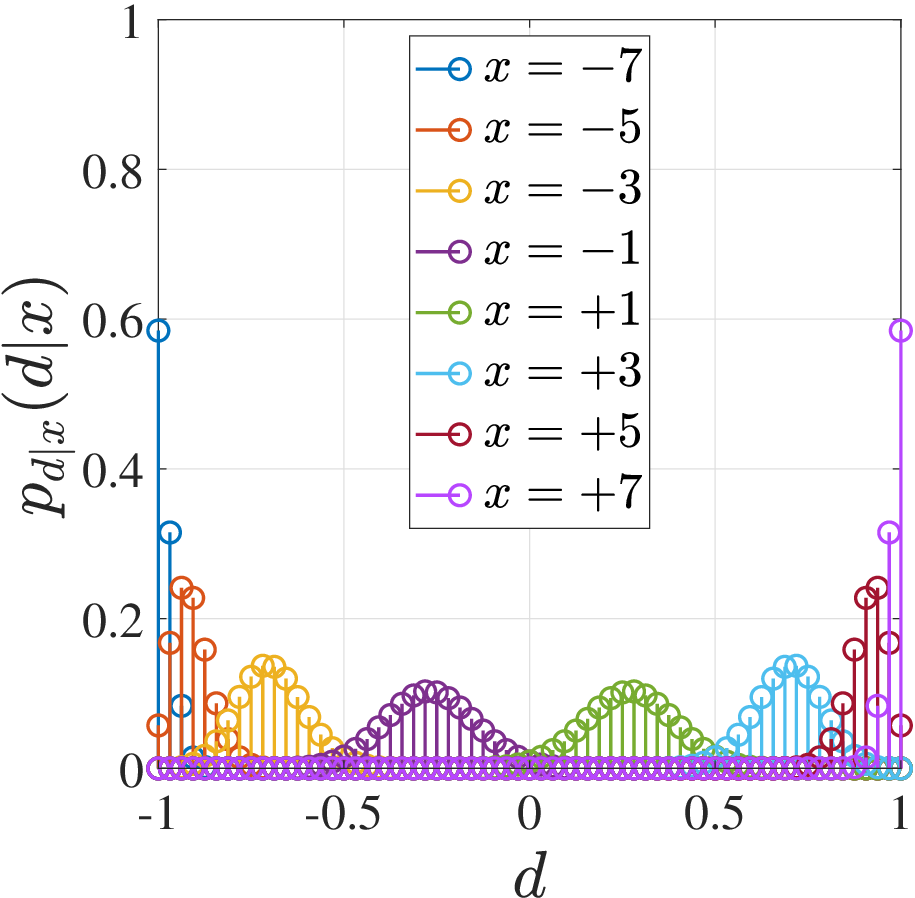}
        \label{fig:d_distr_1bit_8ASK_clas_const_opt_SNR}}
    \hfil
    \subfloat[Optimized 64-\acs{QAM}.]{\includegraphics[width=\plotwidthh\textwidth]{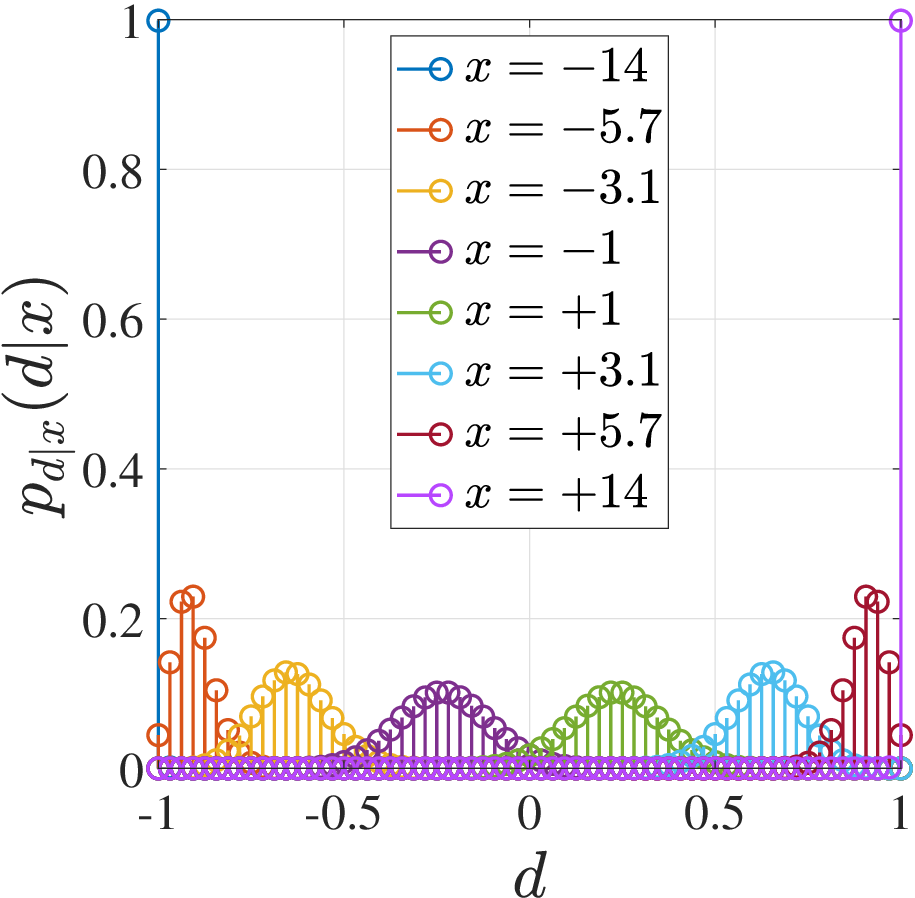}
        \label{fig:d_distr_1bit_8ASK_opt_const_opt_SNR}}
    \caption{\acs{PMF} $p_{\iq{d}|\iq{x}}(\iq{d}|\iq{x})$~(\ref{eq:d_pmf}) for $b=1$, $N=64$, $\Delta=2$, and different constellations at optimum \acs{SNR}, respectively.}
    \label{fig:d_distr_1bit_4ASK_8ASK_const_comparison_opt_SNR}
\end{figure}

In the following, we motivate the reduction of \ac{SER} for the optimized constellations with 1-bit quantization.
For 16-, \mbox{36-,} and 64-\ac{QAM} transmission, we observed that the minimum \ac{SER} can be decreased by increasing the distance of the outermost constellation points to the remaining inner constellation points and simultaneously keeping the inner points almost equidistantly.
This behavior can be explained by comparing the \acp{PMF} $p_{\iq{d}|\iq{x}}(\iq{d}|\iq{x})$~(\ref{eq:d_pmf}) conditioned on the symbols of the classical constellation and the optimized constellation at the \ac{SNR} which yields the minimum \ac{SER}, respectively, shown in Fig.~\ref{fig:d_distr_1bit_4ASK_8ASK_const_comparison_opt_SNR}.
For 16-\ac{QAM}, the main difference between the two \acp{PMF} is that for the optimized constellation (Fig.~\ref{fig:d_distr_1bit_4ASK_opt_const_opt_SNR}) the probability mass for the outer symbols is concentrated in a single detection variable value $\iq{d}=\pm1$, whereas for the classical constellation (Fig.~\ref{fig:d_distr_1bit_4ASK_clas_const_opt_SNR}) the probability mass for the outer symbols is spread over essentially three detection variable values.
Therefore, the overlap between the probability mass for the outer constellation point and the probability mass for the inner constellation point is smaller for the optimized constellation, and, hence, the \ac{SER} is reduced.
The fact that the probability mass for the outer symbols is concentrated at $\iq{d}=\pm1$ means that these symbols almost always fall into the same quantization region of the {\ac{ADC}}, respectively.
The same holds for 64-\ac{QAM} transmission, cf. Figs.~\ref{fig:d_distr_1bit_8ASK_clas_const_opt_SNR} and ~\ref{fig:d_distr_1bit_8ASK_opt_const_opt_SNR}.
In Table~\ref{tab:SER_single_symbols}, the \acp{SER} of single quadrature component constellation symbols are collected.
\begin{table*}[tb]
\caption{Error rates of single quadrature component constellation symbols $\iq{x}\in\mathcal{X}'$ at optimum \ac{SNR} for the \ac{ML} detector~(\ref{eq:ML_det_exact}) with $b=1$ and $N=64$.}
\begin{center}
\begin{tabular}{|c|c|c|c|c|c|c|c|c|}
\multicolumn{9}{c}{Classical constellation} \\
\hline
$\iq{x}\in\mathcal{X}'$&$-7$&$-5$&$-3$&$-1$&$+1$&$+3$&$+5$&$+7$ \\
\hline
16-\acs{QAM}&/&/&$3.9\cdot10^{-4}$&$6.0\cdot10^{-4}$&$6.0\cdot10^{-4}$&$3.9\cdot10^{-4}$&/&/ \\
\hline
36-\acs{QAM}&/&$4.6\cdot10^{-2}$&$7.1\cdot10^{-2}$&$1.3\cdot10^{-2}$&$1.1\cdot10^{-2}$&$7.1\cdot10^{-2}$&$4.6\cdot10^{-2}$&/ \\
\hline
64-\acs{QAM}&$1.0\cdot10^{-1}$&$2.9\cdot10^{-1}$&$9.0\cdot10^{-2}$&$3.7\cdot10^{-2}$&$2.8\cdot10^{-2}$&$9.0\cdot10^{-2}$&$2.9\cdot10^{-1}$&$1.0\cdot10^{-1}$ \\
\hline
\multicolumn{9}{c}{} \\
\multicolumn{9}{c}{Optimized constellation} \\
\hline
$\iq{x}\in\mathcal{X}'$&$-a_3$&$-a_2$&$-a_1$&$-1$&$+1$&$+a_1$&$+a_2$&$+a_3$ \\
\hline
16-\acs{QAM}&/&/&$7.6\cdot10^{-41}$&$5.6\cdot10^{-7}$&$3.3\cdot10^{-7}$&$7.6\cdot10^{-41}$&/&/ \\
\hline
36-\acs{QAM}&/&$7.6\cdot10^{-12}$&$9.2\cdot10^{-3}$&$7.4\cdot10^{-3}$&$5.1\cdot10^{-3}$&$9.2\cdot10^{-3}$&$7.6\cdot10^{-12}$&/ \\
\hline
64-\acs{QAM}&$2.2\cdot10^{-5}$&$7.6\cdot10^{-2}$&$7.4\cdot10^{-2}$&$7.1\cdot10^{-2}$&$5.4\cdot10^{-2}$&$7.4\cdot10^{-2}$&$7.6\cdot10^{-2}$&$2.2\cdot10^{-5}$ \\
\hline
\end{tabular}
\label{tab:SER_single_symbols}
\end{center}
\end{table*}
It can be observed that for a given classical constellation the error rates of all symbols are of same order of magnitude.
However, this is not true anymore for the optimized constellations.
Here, the error rate of the outer constellation points is negligible which is in accordance with the previously presented \acp{PMF}.
Hence, \ac{SER} is determined by the remaining inner symbols.
Also, the reduction in error rate compared to the classical constellation is smaller for the inner symbols than for the outer symbols.
This explains why optimizing the constellation yields a higher gain for 16-\ac{QAM} than for 64-\ac{QAM}.

\subsection{2-Bit Quantization}\label{subsec:2bit_const_opt}
\begin{figure}[t]
    \centerline{\includegraphics[width=\plotwidth\textwidth]{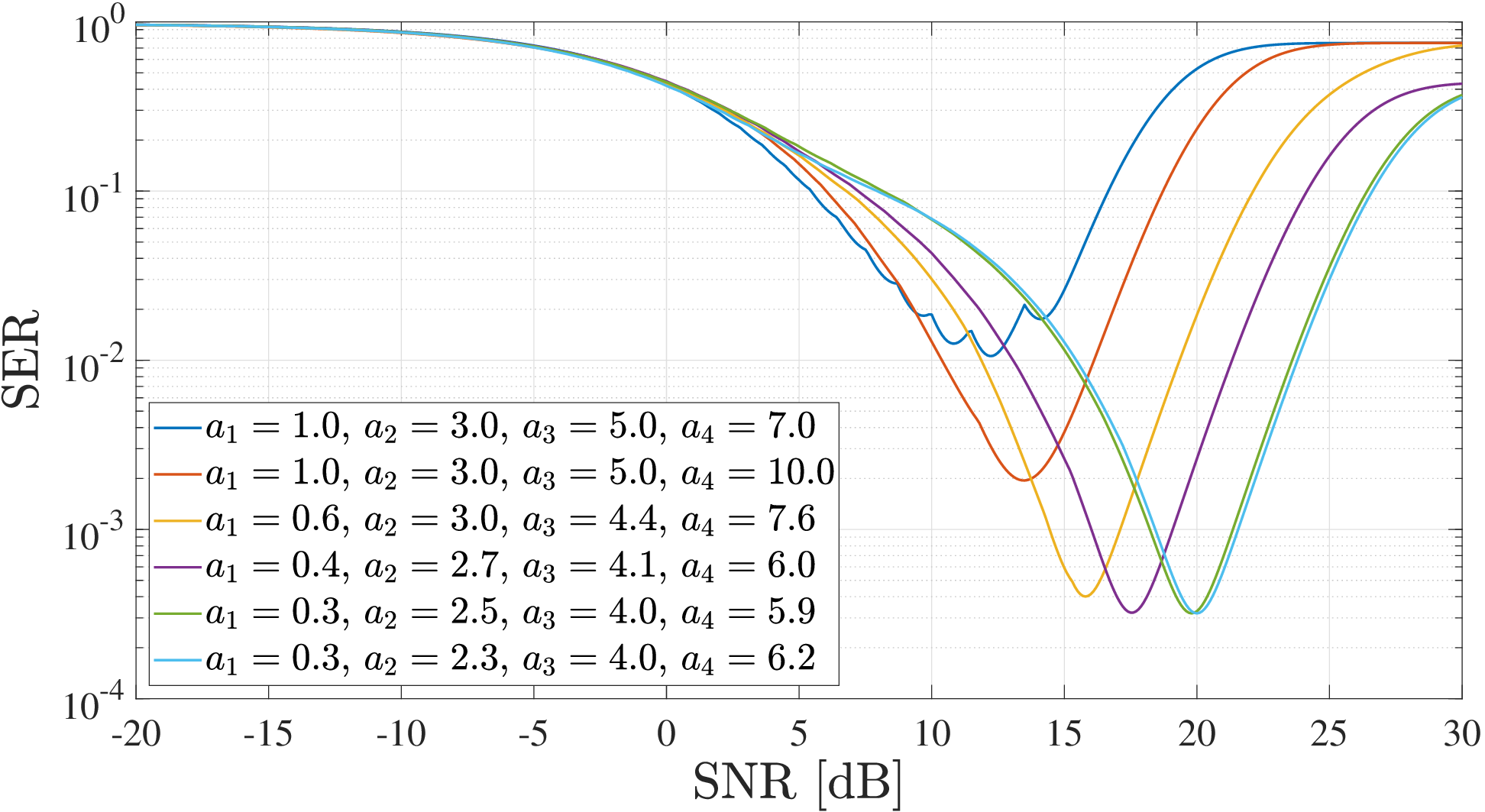}}
    \caption{Analytical {\acs{SER}} for the {\acs{ML}} detector~({\ref{eq:ML_det_exact}}) for different 64-{\ac{QAM}} constellations with $\mathcal{X}'=\{\pm a_1,\pm a_2,\pm a_3,\pm a_4\}$, $b=2$, and $N=32$.}
    \label{fig:SER_2-bit_N32_8-ASK_ML_opt_const}
\end{figure}

\begin{figure}[t]
    \centering
    \subfloat[Classical 64-\acs{QAM}.]{\includegraphics[width=\plotwidthh\textwidth]{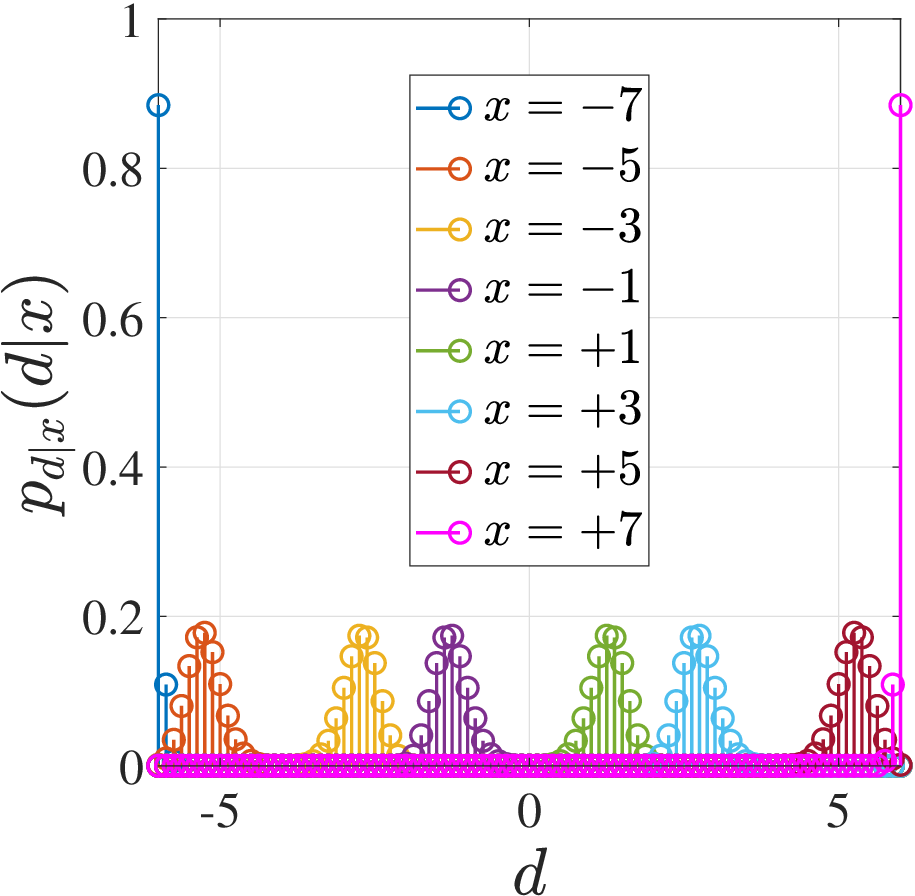}
        \label{fig:d_distr_2bit_8ASK_clas_const_opt_SNR}}
    \hfil
    \subfloat[Optimized 64-\acs{QAM}.]{\includegraphics[width=\plotwidthh\textwidth]{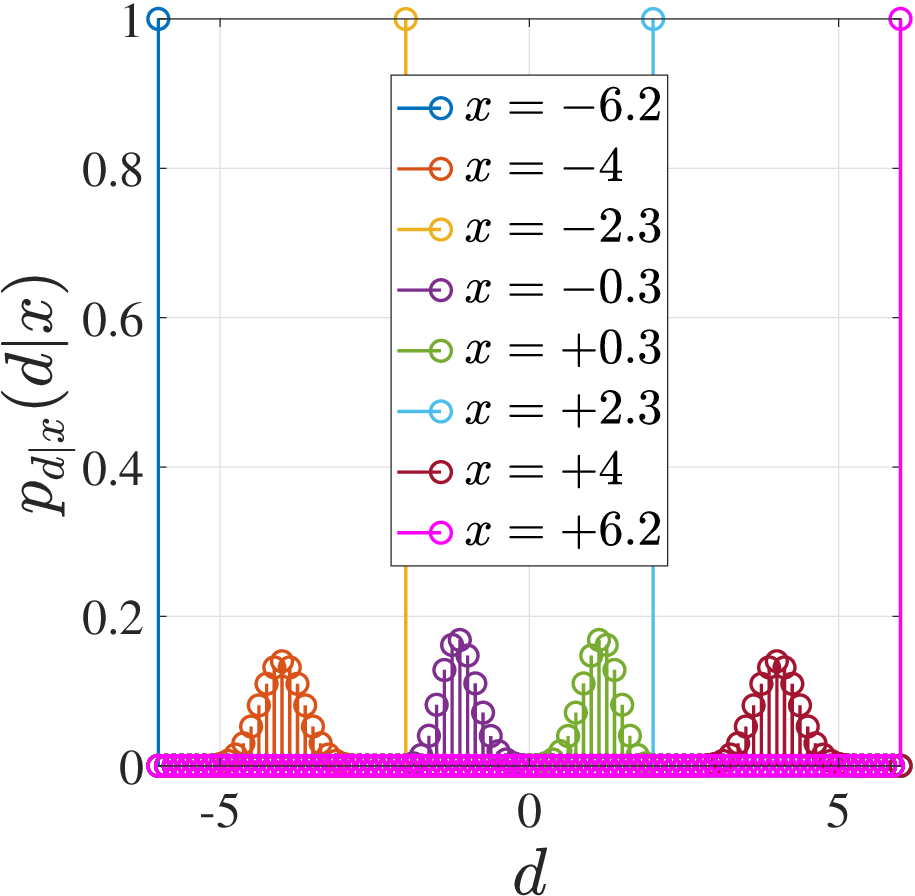}
        \label{fig:d_distr_2bit_8ASK_opt_const_opt_SNR}}
    \caption{{\acs{PMF}} $p_{\iq{d}|\iq{x}}(\iq{d}|\iq{x})$~({\ref{eq:d_pmf}}) for $b=2$, $N=32$, $\Delta=4$, and different constellations at optimum {\acs{SNR}}, respectively.}
    \label{fig:d_distr_2bit_8ASK_const_comparison_opt_SNR}
\end{figure}

Finally, we optimize the 64-{\ac{QAM}} constellation with four degrees of freedom, $\mathcal{X}'=\{\pm a_1,\pm a_2,\pm a_3,\pm a_4\}$, for the case of 2-bit quantization.
We fix the quantization step size to $\Delta=4$ and search for constellations which yield a small minimum {\ac{SER}}.
Since the number of possible constellations is very large in this case, we do not show all corresponding {\ac{SER}} curves.
{\ac{SER}} for selected constellations is depicted in Fig.~{\ref{fig:SER_2-bit_N32_8-ASK_ML_opt_const}}.
Here, we can observe that the minimum achievable {\ac{SER}} can be decreased significantly by using an optimized constellation.
Whereas a classical 64-{\ac{QAM}} constellation with $a_1=1$, $a_2=3$, $a_3=5$, and $a_4=7$ yields a minimum {\ac{SER}} of $1.1\cdot10^{-2}$, with an optimized constellation, a minimum {\ac{SER}} of approximately $3.2\cdot10^{-4}$ is achieved.
Increasing the distance of the outermost constellation points to the remaining inner constellation points as for 1-bit quantization also yields a gain compared to the classical constellation but is not optimal, cf. Fig.~{\ref{fig:SER_2-bit_N32_8-ASK_ML_opt_const}}.
The reason for this will become clear below when considering the {\ac{PMF}}.
Besides, there are again multiple constellations which achieve the minimum {\ac{SER}} at different {\ac{SNR}} values.
There are also multiple constellations which achieve the minimum {\ac{SER}} at the same {\ac{SNR}} due to the additional degree of freedom.
Furthermore, it can be noted that the gain of using an optimized 64-{\ac{QAM}} constellation is much larger for 2-bit quantization than for 1-bit quantization.
The reason for this can be explained via the {\ac{PMF}} $p_{\iq{d}|\iq{x}}(\iq{d}|\iq{x})$~({\ref{eq:d_pmf}}) for different 64-{\ac{QAM}} constellations which is illustrated in Fig.~{\ref{fig:d_distr_2bit_8ASK_const_comparison_opt_SNR}}.
In addition to the probability mass of the outermost symbols of the quadrature component constellation, the probability mass of two inner symbols is concentrated in a single detection variable value $d$ as well for the optimized constellation which minimizes the overlap between the {\acp{PMF}} and, hence, minimizes the overall {\ac{SER}}.
This additional concentration of probability mass is enabled by the two extra quantization regions compared to 1-bit quantization.
In this case, the corresponding receive symbols almost always fall into the same quantization region of the {\ac{ADC}}, i.e., the receive observations for the outer symbols before the {\ac{ADC}} are almost always within $[-2\Delta,-\Delta)$ and $[\Delta,2\Delta]$, respectively, and the observations for the middle inner symbols corresponding to the concentrated probability mass are almost always within $[-\Delta,0)$ and $[0,\Delta)$, respectively.
Due to the additional quantization regions, we are also expecting high gains when optimizing higher-order constellations for $b>2$.

\section{Conclusion}\label{sec:concl}
In this work, we have investigated the \ac{SER} performance of a \ac{THz} band transmission with uniform multi-bit quantization and spatial oversampling at the receiver.
At first, we considered the \ac{ADC} power consumption in general and determined different low-resolution reception schemes with equal \ac{ADC} power consumption for a fair comparison.
Then, the statistics of the detection variable which is generated by combining the quantized observations have been analyzed.
We have formulated the \ac{ML} detector as well as several suboptimal detection schemes.
Our results indicate that the suboptimal detection schemes perform noticeably worse than the optimal \ac{ML} detector when the constellation size is larger than the number of quantization levels of the \ac{ADC}.
We have also shown that 1-bit quantization outperforms 2- and 3-bit quantization at low \acp{SNR} even in case of a transmission with more than \SI{1}{\ac{bpcu}} per real dimension.
Then, we have determined \ac{SER}-optimized 16-, \mbox{36-,} and 64-\ac{QAM} symbol constellations.
For 1-bit quantization, the minimum \ac{SER} can be decreased significantly by increasing the distance of the outermost symbols to the remaining inner symbols.
The effect of this outer point movement is especially strong for small constellations, resulting in a \ac{SER} reduction by several orders of magnitude.
For 2-bit quantization, we can also achieve a large performance gain by optimizing the constellation such that some of the received symbols are falling almost always into a single quantization region of the {\ac{ADC}}.

In future work, we want to generalize our findings on optimum constellations to arbitrary multi-bit quantization and arbitrary oversampling factors and constellation sizes.
Another future topic of interest is channel coding for quantized reception and the exploitation of soft information available due to oversampling in the decoding process.
This can be especially important for low-resolution quantization with higher-order constellations along with a limited number of receive antennas which might exhibit an unsatisfactory performance in practice without coding.

\bibliographystyle{IEEEtran}
\bibliography{IEEEabrv,references}

\end{document}

%% file: abbreviations.tex
\begin{acronym}
\acro{ADC}{analog-to-digital converter}
\acro{ASK}{amplitude-shift keying}
\acro{AWGN}{additive white Gaussian noise}
\acro{bpcu}{bit per channel use}
\acro{CLT}{central limit theorem}
\acro{DAC}{digital-to-analog converter}
\acro{EM}{expectation-maximization}
\acro{ENOB}{effective number of bits}
\acro{iid}[i.i.d.]{independent and identically distributed}
\acro{ISI}{intersymbol interference}
\acro{LoS}{line-of-sight}
\acro{MIMO}{multiple-input multiple-output}
\acro{ML}{maximum-likelihood}
\acro{NLoS}{non-line-of-sight}
\acro{OFDM}{orthogonal frequency division multiplexing}
\acro{PAM}{pulse amplitude modulation}
\acro{PDF}{probability density function}
\acro{PMF}{probability mass function}
\acro{PSK}{phase-shift keying}
\acro{QAM}{quadrature amplitude modulation}
\acro{RF}{radio frequency}
\acro{RRC}{root-raised-cosine}
\acro{SAR}{successive approximate register}
\acro{SER}{symbol error rate}
\acro{SISO}{single-input single-output}
\acro{SIMO}{single-input multiple-output}
\acro{SNDR}{signal-to-noise-and-distortion ratio}
\acro{SNR}{signal-to-noise ratio}
\acro{THz}{Terahertz}
\acro{UPA}{uniform planar array}
\acro{VGA}{variable gain amplifier}
\acro{ZXM}{zero crossing modulation}
\end{acronym}

%% file: macros.tex
\newcommand{\lvec}[1]{\ensuremath{\mathbf{#1}}}  
\newcommand{\gvec}[1]{\ensuremath{\boldsymbol{#1}}}       
\newcommand{\lvecg}[1]{\ensuremath{\mathbf{#1}^\text{c}}}  
\newcommand{\lvecgconj}[1]{\ensuremath{\mathbf{#1}^{\text{c}*}}}  
\newcommand{\lmat}[1]{\ensuremath{\mathbf{#1}}}  
\newcommand{\gmat}[1]{\ensuremath{\boldsymbol{#1}}}       
\newcommand{\er}{\mbox{e}}
\newcommand{\imag}{\ensuremath{\mathrm{j}}}
\newcommand*{\argmin}{\ensuremath{\mathop{\mathrm{arg\,min}}}}
\newcommand*{\argmax}{\ensuremath{\mathop{\mathrm{arg\,max}}}}
\newcommand{\est}[2][]{{E}_{#1}\!\left\{#2\right\}}
\newcommand{\Rset}{\mathbb{R}}
\newcommand{\Nset}{\mathbb{N}}
\newcommand{\diag}[1]{\ensuremath{\mathrm{diag}\!\left\{#1\right\}}}

\newcommand{\iq}[1]{\ensuremath{#1}}
\newcommand{\cmplx}[1]{\ensuremath{#1^\text{c}}}